\documentclass{article}

\usepackage{PRIMEarxiv}
\usepackage{fancyhdr}       

\def\lim{\mathop{\mathsf{lim}}} 

\def\log{\mathrm{log}}

\newcommand{\bA}{\mathbf{A}}
\newcommand{\by}{\mathbf{y}}
\newcommand{\bY}{\mathbf{Y}}
\newcommand{\bx}{\mathbf{x}}

\newcommand{\bW}{\mathbf{W}}
\newcommand{\bz}{\mathbf{z}}

\newcommand{\bw}{\mathbf{w}}
\newcommand{\bu}{\mathbf{u}}
\newcommand{\bn}{\mathbf{n}}
\newcommand{\bs}{\mathbf{s}}

\newcommand{\bS}{\mathbf{S}}
\newcommand{\bQ}{\mathbf{Q}}
\newcommand{\bB}{\mathbf{B}}
\newcommand{\bI}{\mathbf{I}}

\newcommand{\dd}{\mathrm{d}}
\newcommand{\grad}{\nabla}

\usepackage{tikz}
\usepackage{pgfplots}
\usetikzlibrary{spy,calc}

\newif\ifblackandwhitecycle
\gdef\patternnumber{0}

\pgfkeys{/tikz/.cd,
    zoombox paths/.style={
        draw=green,
        very thick
    },
    black and white/.is choice,
    black and white/.default=static,
    black and white/static/.style={ 
        draw=white,   
        zoombox paths/.append style={
            draw=white,
            postaction={
                draw=black,
                loosely dashed
            }
        }
    },
    black and white/static/.code={
        \gdef\patternnumber{1}
    },
    black and white/cycle/.code={
        \blackandwhitecycletrue
        \gdef\patternnumber{1}
    },
    black and white pattern/.is choice,
    black and white pattern/0/.style={},
    black and white pattern/1/.style={    
        draw=white,
        postaction={
            draw=black,
            dash pattern=on 2pt off 2pt
        }
    },
    black and white pattern/2/.style={    
        draw=white,
        postaction={
            draw=black,
            dash pattern=on 4pt off 4pt
        }
    },
    black and white pattern/3/.style={    
        draw=white,
        postaction={
            draw=black,
            dash pattern=on 4pt off 4pt on 1pt off 4pt
        }
    },
    black and white pattern/4/.style={    
        draw=white,
        postaction={
            draw=black,
            dash pattern=on 4pt off 2pt on 2pt off 2pt on 2pt off 2pt
        }
    },
    zoomboxarray inner gap/.initial=5pt,
    zoomboxarray columns/.initial=1,
    zoomboxarray rows/.initial=1,
    figurename/.initial={zoombox},
    zoomboxarray/.style={
        execute at begin picture={
            \begin{scope}[
                spy using outlines={%
                    zoombox paths,
                    width=0.4*\imagewidth,  
                    height=0.4*\imageheight, 
                    magnification=3,        
                    every spy on node/.style={
                        zoombox paths
                    },
                    every spy in node/.style={
                        zoombox paths
                    }
                }
            ]
        },
        execute at end picture={
            \end{scope}
            \gdef\patternnumber{0}
        },
        spymargin/.initial=0.1em,
        zoomboxes xshift/.initial=0,
        zoomboxes yshift/.initial=0,
        caption margin/.initial=4ex,
    },
    adjust caption spacing/.code={},
    image container/.style={
        inner sep=0pt,
        at=(image.north),
        anchor=north,
        adjust caption spacing
    },
    zoomboxes container/.style={
        inner sep=0pt,
        at=(image.south),
        anchor=north,
        name=zoomboxes container,
        adjust caption spacing
    },
    calculate dimensions/.code={
        \pgfpointdiff{\pgfpointanchor{image}{south west} }{\pgfpointanchor{image}{north east} }
        \pgfgetlastxy{\imagewidth}{\imageheight}
        \global\let\imagewidth=\imagewidth
        \global\let\imageheight=\imageheight
        \gdef\columncount{1}
        \gdef\rowcount{1}
        
    },
    image node/.style={
        inner sep=0pt,
        name=image,
        anchor=south west,
        append after command={
            [calculate dimensions]
        }
    },
    color code/.style={
        zoombox paths/.append style={draw=#1}
    },
    connect zoomboxes/.style={
        spy connection path={\draw[draw=none,zoombox paths] (tikzspyonnode) -- (tikzspyinnode);}
    },
    help grid code/.code={
        \begin{scope}[
            x={(image.south east)},
            y={(image.north west)},
            font=\footnotesize,
            help lines,
            overlay
        ]
        \foreach \x in {0,1,...,9} { 
            \draw(\x/10,0) -- (\x/10,1);
            \node [anchor=north] at (\x/10,0) {0.\x};
        }
        \foreach \y in {0,1,...,9} {
            \draw(0,\y/10) -- (1,\y/10);                        
            \node [anchor=east] at (0,\y/10) {0.\y};
        }
        \end{scope}    
    },
    help grid/.style={
        append after command={
            [help grid code]
        }
    },
}

\newcommand\zoombox[2][]{
    \begin{scope}[zoombox paths]
        \pgfmathsetmacro\xpos{
            0 
        }
        \pgfmathsetmacro\ypos{
            0 
        }
        \edef\dospy{\noexpand\spy [
            #1,
            zoombox paths/.append style={
                black and white pattern=\patternnumber
            },
            every spy on node/.append style={#1},
            x=\imagewidth,
            y=\imageheight
        ] on (#2) in node [anchor=south west] at ($(image.south west)+(\xpos pt,\ypos pt)+(1 pt,1 pt)$);}
        \dospy
        \pgfmathtruncatemacro\pgfmathresult{ifthenelse(\columncount==\pgfkeysvalueof{/tikz/zoomboxarray columns},\rowcount+1,\rowcount)}
        \global\let\rowcount=\pgfmathresult
        \pgfmathtruncatemacro\pgfmathresult{ifthenelse(\columncount==\pgfkeysvalueof{/tikz/zoomboxarray columns},1,\columncount+1)}
        \global\let\columncount=\pgfmathresult
        \ifblackandwhitecycle
            \pgfmathtruncatemacro{\newpatternnumber}{\patternnumber+1}
            \global\edef\patternnumber{\newpatternnumber}
        \fi
    \end{scope}
}

\usepackage{array}
\usepackage[caption=false,font=normalsize,labelfont=sf,textfont=sf]{subfig}
\usepackage{textcomp}
\usepackage{stfloats}
\usepackage{textcomp}
\usepackage{verbatim}
\usepackage{hyperref}       
\usepackage{tabularx}
\usepackage{booktabs}       
\usepackage{nicefrac}       
\usepackage{microtype}      
\usepackage{bm}				
\usepackage{graphicx}		
\usepackage{mathtools}		
\usepackage{algorithm}
\usepackage{algcompatible}
\usepackage{nicefrac}
\usepackage{comment}
\usepackage{enumitem}		
\usepackage{multirow}       
\usepackage{tabularx}	    
\usepackage{hhline}
\usepackage{arydshln}		
\usepackage{xcolor}
\usepackage{mathtools,nccmath}
\usepackage{amsmath}
\newcommand{\triangleq}{\mathrel{\overset{\triangle}{=}}}
\usepackage{tikz}
\usepackage{pgfplots}
\pgfplotsset{compat=1.18}
\usetikzlibrary{spy,calc}
\definecolor{lightgreen}{rgb}{.9,1,.9}
\definecolor{red}{rgb}{1,0,0}

\newcolumntype{L}[1]{>{\raggedright\arraybackslash}p{#1}}
\newcolumntype{C}[1]{>{\centering\arraybackslash}p{#1}}
\newcolumntype{R}[1]{>{\raggedleft\arraybackslash}p{#1}}

\usepackage{amsfonts}
\usepackage{graphicx}
\usepackage{epstopdf}
\ifpdf
  \DeclareGraphicsExtensions{.eps,.pdf,.png,.jpg}
\else
  \DeclareGraphicsExtensions{.eps}
\fi

\usepackage[T1]{fontenc}
\usepackage{lmodern}

\usepackage{epstopdf}

\ifpdf
\hypersetup{
  pdftitle={Bridging data-driven priors via the score function for posterior sampling -- Comparative review and experimental study},
  pdfauthor={E. C. Faye, M. D. Fall, S. Delchini and N. Dobigeon}
}
\fi

\title{Bridging data-driven priors via the score function for posterior sampling -- Comparative review and experimental study}

\author{
    \hspace{0.5cm}
    \And
  Elhadji C. Faye   \\
    IDP, Univ Orl\'eans\\
    Orl\'eans, France\\
   \And 
   \hspace{0.5cm}  
   \And 
  Mame Diarra Fall\thanks{Part of this work was supported by the AI.iO Project (ANR-20-THIA-0017) and the BACKUP project (ANR-23-CE40-0018-01).}   \\
    LITIS, Univ Rouen Normandie\\
    Rouen, France\\
    \hspace{0.5cm}
   \And
   	Sylvain Delchini\\
   	Bureau de Recherches Géologiques et Minières\\
   	Orléans, France
     \And
   Nicolas Dobigeon\thanks{Part of this work was supported by the Artificial and Natural Intelligence Toulouse Institute (ANITI), funded by the France 2030 program under the grant agreement ANR-23-IACL-0002.}$\ $\thanks{Corresponding author: \texttt{Nicolas.Dobigeon@irit.fr}.}\\
    IRIT, Univ Toulouse \\
  Toulouse, France
}

\begin{document}
\maketitle

\begin{abstract}
This paper reviews how a diverse set of popular data-driven priors commonly used in Bayesian inverse problems can be unified through their respective score functions. By framing these priors under this common perspective, we show that they can benefit from their straightfoward and effective integration into a recently proposed sampling algorithm. The applicability of this common framework is illustrated by considering several data-driven priors, namely regularization-by-denoising, normalizing flow-based priors, score-based generative models, and convex-ridge regularizers. For these four particular priors, the performance of the method is evaluated when conducting image inpainting and single image super-resolution. These results, as well as those obtained when restoring real images acquired in a geological context, demonstrate the efficiency of the method. This unified framework proves versatile enough to handle any posterior distribution defined by a broad class of score function-based priors, beyond the specific cases considered in this paper.
\end{abstract}


\newpage

\section{Introduction}\label{sec:intro}
Many image restoration (IR) tasks can be cast as linear inverse problems characterized by the forward model 
\begin{equation}\label{eq:forward_model}
 \by = \bA \bx + \bn   
\end{equation}
where $\by$ denote the measurements, $\bA$ is the degradation matrix and $\bn$ is assumed to be an additive white Gaussian noise \cite{Marchall2023,Sun2023}. The goal is then to recover the unknown image $\bx \in \mathbb{R}^n$ from the degraded version $\by \in \mathbb{R}^m$. Notable examples include super-resolution, denoising, deblurring, as well as inpainting. Within a statistical framework, the relationship between $\bx$ and $\by$ is described by the likelihood function 
\begin{equation}\label{eq:likelihood}
p(\by|\bx) \propto \exp\left[-f(\bx,\by)\right]    
\end{equation}
where 
\begin{equation}\label{eq:potential}
f(\bx, \by)= \tfrac{1}{2\sigma^2}\|\bA \bx - \by \|^2_2
\end{equation}
is the data fidelity term. Since IR is typically an ill-posed or, at least, an ill-conditioned problem, the Bayesian framework is known to be well suited to explicitly include regularizations by assigning a distribution to $\bx$ summarizing any prior knowledge on the unknown restored image. This prior distribution takes the form
\begin{equation}
p(\bx) \propto \exp\left[-g(\bx)\right]    
\end{equation}
where $g(\cdot)$ denotes the regularization potential generally chosen to promote an expected particular feature exhibited by the sought solution. Given the likelihood function and the prior, the posterior distribution $p(\bx|\by)$ writes
\begin{equation}
    p(\bx|\by) \propto  \exp \left[ -f(\bx,\by)- g(\bx) \right].
    \label{posterior}
\end{equation}
A straightforward way to  exploit this posterior distribution boils down to derive the maximum a posteriori (MAP) estimator which is obtained by solving the  minimization problem
\begin{equation}
\hat{\bx} = \operatornamewithlimits{arg   min}_{\bx} f(\bx,\by) + \beta g(\bx).
\label{eq:MAP}
\end{equation}
Solving the optimization problem \eqref{eq:MAP} typically relies on iterative gradient-like methods, as closed-form solutions are rarely available except in simpler cases, e.g.,  where both the data-fidelity and regularization terms are quadratic. Advances in convex optimization have enabled efficient resolutions of such high-dimensional problems, even when the objective function is not smooth. Among the commonly used iterative schemes frequently used in imaging problems are the alternating direction method of multipliers (ADMM) \cite{afonso2010augmented} and the first-order primal-dual algorithm \cite{chambolle2011first}.

However, the resolution framework discussed above is able to only provide point estimates. An alternative approach consists in investigating the full posterior distribution \eqref{posterior}, which provides a comprehensive probabilistic description of the space of the solutions. This approach allows one not only to derive other Bayesian estimators such as the minimum mean square error (MMSE) estimator $\hat{\bx}_{\mathrm{MMSE}} = \mathbb{E}[ \bx|\by]$ but also to perform uncertainty quantification by computing credibility intervals \cite{bardsley2018computational,cai2018uncertainty}. However evaluating such Bayesian quantities require to compute integrals, a computationally infeasible task in high dimensions. To address this issue, Markov chain Monte Carlo (MCMC) algorithms can be employed to approximate these integrals through sampling. Developing efficient MCMC algorithms for Bayesian imaging may face two major obstacles: the high dimensionality of the problem and the non-smoothness of the regularization function $g(\cdot)$. To tackle these challenges, some recent works have capitalized on advanced tools borrowed from the non-smooth convex optimization literature. For instance, the popular Moreau Yosida unadjusted Langevin algorithm (MYULA) introduces an unadjusted Langevin algorithm (ULA) which targets a smooth approximation of the posterior \cite{durmus2018efficient}. To sample from distributions defined by non-differentiable potentials, alternatives consist in resorting to other  proximal sampling algorithms \cite{lau2022non}, subgradient methods \cite{habring2024subgradient} or stochastic counterparts of primal-dual algorithms \cite{durmus2019analysis,salim2020primal,narnhofer2024posterior}. Yet, these MCMC methods have often been considered to sample from posterior distributions defined by hand-crafted model-based priors, such as total variation (TV) \cite{rudin1992nonlinear,pereyra2016proximal,Sun2023}, Laplacian-based smoothing priors \cite{orieux2012sampling,vono2019split} or sparsity-inducing priors \cite{lee2010hierarchical, park2008bayesian, dobigeon2009hierarchical}. Unfortunately, in the context of imaging problems, designing appropriate model-based regularizations remains empirical, and might fail to capture sophisticated structures inherent to natural images effectively. 

With the advent of deep learning (DL), recent works have explored the benefit of using data-driven regularizations \cite{arridge2019solving}. Instead of relying on handcrafted regularizations, the prior distribution $p(\bx)$ is rather learned from available data sets and encoded into a neural network (NN). Following a synthesis regularization paradigm adopted by Bora \emph{et al.} \cite{bora2017compressed} in a deterministic framework, a large family of methods designs this prior model through a generative model $\bx=\mathsf{G}(\bz)$ where the latent variable $\bz$ is typically drawn from a simple instrumental distribution, $\bz \sim \mathcal{N}(\boldsymbol{0}, \mathbf{I})$. Once trained, a generative model is able to produce realistic data samples by encoding the mapping from the lower-dimensional (Gaussian) distribution of the latent variable $\bz$ towards the distribution of the training samples in the original space. After deriving the posterior distribution of the latent variable, inference can then be conducted in the latent space using dedicated  sampling-based methods \cite{holden2022bayesian}. When the generative model is parametrized by an external quantity $\mu$, the prior distribution writes $p(\bx|\mu)$ and is defined through a conditional generative prior \cite{melidonis2024empirical}. The joint estimation of the latent variable $\bz$ and the hyperparameter $\mu$ require to implement specific numerical methods, such as the stochastic optimization unadjusted Langevin \cite{debortoli2021efficient}. Conversely, these generative models can be also exploited to conduct the inference   in the original high-dimensional space directly, as already investigated in \cite{oberlin2021regularization} in a deterministic framework under the analysis regularization paradigm. To circumvent the resulting computational difficulty, the authors in \cite{Marchall2023} derive a  Laplace approximation of the generative model, yielding an iterative minimization algorithm. It is worth noting that most of the aforementioned methods mainly rely on (variational) autoencoders (VAEs) or generative adversarial networks (GANs) as generative models, known to be impaired by their limited expressiveness when compared to more advanced models. Alternatives consist in resorting to more recent neural network-based models such as normalizing flows (NF), which have also the great advantage of providing an explicit change of variables from the latent to the original spaces. Capitalizing on the closed-form expression of the prior distribution, Cai \emph{et al.}  propose to sample from the resulting posterior distribution directly in the image domain \cite{cai2023nf}, even if sampling in the latent domain may avoid pathological behaviors when targeting complex distributions \cite{Coeurdoux_MLJ_2024}. Following a different strategy, instead of building on generative models to design explicit prior models, deep NN can be embedded into Monte Carlo sampling schemes as implicit regularizers. One archetypal instance of such kind of approaches is the plug-and-play ULA (PnP-ULA) which leverages an implicit prior through the use of a deterministic denoiser \cite{laumont2022bayesian}. Interestingly, while following distinct motivations and adopting different data-driven priors, these Bayesian IR resolutions share two key features. Firstly, most of the aforementioned sampling methods benefit from the possibility of explicitly computing the so-called prior score function, i.e., the gradient of the log-prior distribution. Secondly, they basically rely on an rather simple ULA scheme, that could show limited performance when tackling severely ill-conditioned problems. 

This paper proposes to build on these latter observations by revisiting Bayesian IR from a unified perspective -- provided that the prior score function is explicit or can be efficiently evaluated -- while going beyond the limitation of a simple ULA sampling scheme. Specifically, one shows that a sampling algorithm, recently proposed in the literature and referred to as  Langevin-within-SGS (LwSGS), is sufficiently versatile to efficiently sample from posterior distributions derived from  any score function-based priors.  Adopting an  asymptotically augmented data augmentation (AXDA) \cite{vono2020asymptotically}, LwSGS is defined as a variant of the split Gibbs sampler (SGS) \cite{vono2019split}. SGS follows a \emph{divide-and-conquer} strategy  to decompose the initial sampling problem into several simpler individual sampling tasks. It has shown to efficiently scale in high dimensions and to significantly improve the mixing properties of the produced Markov chains \cite{vono2019split}. Inheriting from these key features, LwSGS additionally embeds a Langevin Monte Carlo step which is particularly well suited to handle the score function-based regularizations considered in this work. Among the class of data-driven priors dealt by the proposed framework, archetypal examples considered in this paper include the already aforementioned models, namely the PnP priors coined as regularization-by-denoising (RED) \cite{laumont2022bayesian} and NF-based priors \cite{cai2023nf}. At no cost, it is shown to be able to straightforwardly deal also with score-based generative models (SGM)  and convex-ridge regularizers (CRR)  \cite{goujon2023neural}. 

The sequel of this paper is organized as follows. Section \ref{sec:score} provides some background on the score function. Section \ref{sec:LwSGS} recalls the LwSGS algorithm and highlights its versatility when handling data-driven priors defined by score functions. Besides, this section also discusses four particular instances of such priors, namely RED, SGM, NF and CRR, that are envisioned from this score function perspective. In Section \ref{sec:experiments}, an experimental study is conducted on simulated data sets to compare the performance of the resulting instances of LwSGS to their ULA-based counterparts when performing image inpainting and superresolution. Section \ref{sec:real_data} illustrates the practicability of the method when restoring real-world imaging data of geological tailing samples. Finally,  Section \ref{sec:conclusion} concludes this paper.

\section{Score function: background}\label{sec:score}

Let us consider a probability density function $p(\bx;\boldsymbol{\theta})$ associated with the variable $\bx$ and defined by parameters $\boldsymbol{\theta}$. In agreement with \cite{hyvarinen2005estimation}, this paper defines the so-called score function $\bs(\bx;\boldsymbol{\theta})$ as the gradient of the log-density with respect to the variable $\bx$, i.e.,
\begin{equation}
    \bs(\bx;\boldsymbol{\theta}) \triangleq \nabla_{\bx} \log\, p(\bx;\boldsymbol{\theta}).
\end{equation}
It is worth noting that this terminology departs from the alternate definition of the \emph{score} conventionally admitted in the statistics literature, where it generally  refers to the gradient of the log-density with respect to the parameters $\boldsymbol{\theta}$ \cite[Chap. 4]{serfling2009approximation}. This score function has shown to be a crucial quantity to perform various tasks in statistical learning, in particular density estimation. In a nutshell, suppose that one wants to approximate a target distribution denoted as $\pi(\bx)$ with an instrumental distribution $p(\bx; \boldsymbol{\theta})$. Conventional variational inference adjusts the parameters $\boldsymbol{\theta}$ to minimize the Kullback divergence between the target and instrumental distributions. An efficient alternative, particularly appealing when dealing with non-normalized distributions, consists in searching for the parameters $\boldsymbol{\theta}$ which rather minimize the Fisher divergence, i.e., the mean square error between the respective score functions of the target and instrumental distributions. The so-called \emph{score matching} estimator is thus defined as
\begin{equation} \label{eq:score_matching}
    \hat{\boldsymbol{\theta}} = \operatornamewithlimits{argmin}_{\theta} \frac{1}{2} \int \pi(\bx) \left\| \bs(\bx;\boldsymbol{\theta})  - \nabla_{\bx} \log\, \pi(\bx) \right\|^2 d\bx.
\end{equation}
When the target distribution $\pi(\bx)$ is not known but only defined through samples, the empirical counterpart of this strategy initially appears as a simple yet efficient alternative to the maximum likelihood estimation \cite{hyvarinen2005estimation,hyvarinen2007some}. Later, it has shown to have much more profound implications \cite{Lyu2009uai}, in particular since its connection to denoising autoencoders drawn in \cite{vincent2011connection} and its decisive role when training score-based generative models \cite{song2019generative}. 

This work  elaborates on this definition of the score function, beyond the restrictive scope of generative modeling. In particular, this key yet common quantity naturally arises when sampling according to posterior distributions derived from various data-driven prior models $p(\bx)$. When possible in the sequel of the paper, to lighten the notations but without any possible ambiguity, these prior score functions will be denoted $\bs(\bx) = \nabla \log\, p(\bx)$, i.e., after omitting the parameters $\boldsymbol{\theta}$ when referring to the score function $\bs(\bx)$ and after omitting the variable $\bx$ when computing the gradient $\nabla  \log\, p(\bx) $.

\section{Sampling algorithm with score function-based priors}\label{sec:LwSGS}
\subsection{Langevin-within-Split Gibbs Sampler}
This section reviews a systematic yet efficient algorithm recently proposed in the literature. While initially implemented in  \cite{Faye2024} to handle a RED prior, we highlight hereafter that this algorithm is sufficiently versatile to sample from any posterior of the form \eqref{posterior} once the prior distribution has a score function which is explicit or easily computable. This algorithm leverages an asymptotically exact data augmentation  \cite{vono2020asymptotically} which has shown to lead to efficient and flexible sampling schemes in various contexts, for instance offering the possibility of embedding deep generative model-based stochastic denoisers as priors, similar to a PnP fashion \cite{bouman2023generative, Coeurdoux2024pnp, wu2024principled}. Starting from the posterior distribution \eqref{posterior}, an auxiliary variable $\bz \in \mathbb{R}^n$ is introduced to form the augmented posterior distribution given by
    \begin{eqnarray}\label{split_dist}
       \pi_{\rho}(\bx, \bz) &=& p({\bx}, {\bz}|{\by}; \rho^2) \nonumber \\
       &\propto& \exp \left[ - f(\bx,\by) - g(\bz) - \tfrac{1}{2\rho^2}|| \bx - \bz||^2 \right]
    \end{eqnarray}
where  $\rho > 0 $ is a parameter controlling  the dissimilarity between $\bx$ and $\bz$. It can be shown that the marginal posterior distribution
$$\pi_{\rho}(\bx) = \int_{\mathbb{R}^n} \pi_{\rho}(\bx, \bz) \dd \bz$$ converges in total variation to the target posterior distribution $p(\bx|\by)$ as $\rho \rightarrow 0$ \cite{vono2019split}. SGS then alternatively samples from the two corresponding conditional distributions associated to the augmented posterior $\pi_{\rho}(\bx, \bz)$
\begin{align}
 p(\bx|\by,\bz;\rho^2 ) &\propto \exp \left[ -  f(\bx,\by) - \tfrac{1}{2\rho^2}|| \bx - \bz||^2 \right]
    \label{eq:condx} \\
    p(\bz | \bx;\rho^2) &\propto \exp \left[ - g(\bz) - \tfrac{1}{2\rho^2}||\bx - \bz||^2 \right].
    \label{eq:condz}
\end{align}
This separation enables the two components, $f(\cdot,\by)$ and $g(\cdot)$, which together define the overall potential, to be disentangled and incorporated into two separate conditional distributions. As a result, SGS benefits from the well-known advantages already demonstrated by its deterministic counterparts such as half quadratic splitting, e.g., simpler implementations. When considering linear Gaussian inverse problems defined by the quadratic potential \eqref{eq:potential}, the conditional distribution $p(\bx|\by, \bz;\rho^2)$ writes 
\begin{equation}\label{gaussian_step}
    p(\bx|\by, \bz;\rho^2) = \mathcal{N}(\bx ; \boldsymbol{\mu}(\bz), \bQ^{-1})
\end{equation} 
where the precision matrix $\bQ$ and the mean vector $\boldsymbol{\mu}(\cdot)$ are
\begin{equation}\label{eq:def_mu_and_Q}
    \left \{
   \begin{array}{r c l}
      \bQ &=& \tfrac{1}{\sigma^2} \bA^\top \bA + \tfrac{1}{\rho^2} \bI \\
      \boldsymbol{\mu}(\bz) &=& \bQ^{-1} \left( \tfrac{1}{\sigma^2} \bA^\top \by + \tfrac{1}{\rho^2} \bz  \right).
   \end{array}
   \right.
\end{equation}
In this case, efficient sampling from this conditional distribution can be accomplished using dedicated algorithms tailored to the structure of the precision matrix $\bQ$ \cite{vono2022high}.

Regarding the second step of the SGS, when the gradient of the log-prior (i.e, the prior score function) is explicit or can be easily computed or approximated, sampling from \eqref{eq:condz} can be conducted by using a Langevin Monte Carlo (LMC) step, i.e., 
\begin{align}\label{LMC_step}
    \bz^{(t+1)} &= \bz^{(t)} + \gamma \nabla \log\, p\left(\bz^{(t)} \mid \bx ;\rho^2\right) + \sqrt{2\gamma} \boldsymbol{\varepsilon}^{(t)}
\end{align}
where $\gamma > 0$ is a fixed step-size and $\boldsymbol{\varepsilon}^{(t)}$ is an $n$-dimensional standard Gaussian random variable. A striking benefit of employing this splitting technique is the following: contrary to ULA-based schemes implemented for instance by \cite{laumont2022bayesian}, \cite{durmus2018efficient} and \cite{cai2023nf} which target the posterior distribution directly, the step size $\gamma$ is here independent of the Lipschitz constant related to the likelihood function, that could be dramatically big for severely ill-conditioned inverse problems. Besides, the discretization in \eqref{LMC_step} of the underlying stochastic differential equation is known to generate a biased Markov chain  that can be corrected by introducing  a Metropolis-Hastings step, at an additional computational cost \cite{dwivedi2018log}. For computational efficiency purpose, as also advocated in \cite{durmus2018efficient} \cite{laumont2022bayesian} and \cite{cai2023nf}, this correction will not be implemented.

Interestingly, given the particular form of the conditional distribution \eqref{eq:condz}, the gradient of the corresponding log-density can be rewritten using the prior score
\begin{align}\label{LMC_step2}
    \nabla \log\, p\left(\bz^{(t)}\mid\bx ;\rho^2\right) &= - \nabla g\left(\bz^{(t)}\right) + \tfrac{1}{\rho^2} \left(\bx - \bz^{(t)}\right) \nonumber  \\
    &= \underbrace{ \nabla \log \, p\left(\bz^{(t)}\right)}_{\text{score function}} + \tfrac{1}{\rho^2} \left(\bx - \bz^{(t)}\right) \nonumber  \\
    & = \bs\left(\bz^{(t)}\right) + \tfrac{1}{\rho^2} \left(\bx - \bz^{(t)}\right).
\end{align}

The convergence of the proposed algorithm, coined as Langevin-within-SGS (LwSGS) and summarized in Algorithm~\ref{alg:LwSGS}, has been rigorously analyzed and extensively discussed in  \cite{Faye2024} when the prior derives from the RED paradigm. However, beyond this special case, it clearly appears that LwSGS is sufficiently generic to be implemented whenever the posterior of interest derived from a differentiable log-prior with an explicit score function. This prerequisite covers several topical cases widely documented in the literature and detailed in what follows.

\newcommand{\algocomment}[1]{ \STATEx {\color[rgb]{0.5,0.8,0.5}{\% \textit{#1}}}}
\begin{algorithm}[H]
\caption{Langevin-within Split Gibbs Sampler (LwSGS)}
\label{alg:LwSGS}
\begin{algorithmic}[1]
\renewcommand{\algorithmicrequire}{\textbf{Input:}}
\REQUIRE prior score function $\bs(\cdot)$, coupling parameter $\rho^2$, step-size $\gamma$, number of burn-in iterations $N_{\mathrm{bi}}$, total number of iterations $N_{\mathrm{MC}}$
\renewcommand{\algorithmicrequire}{\textbf{Initialization:}}  
\REQUIRE $\bx^{(0)}$, $\bz^{(0)}$
\FOR{$t = 0$ to $N_{\mathrm{MC}}-1$}
    \algocomment{Sampling the splitting variable according to \eqref{LMC_step}}
    \STATE $\boldsymbol{\varepsilon}^{(t)} \sim \mathcal{N}(\boldsymbol{0},\mathbf{I})$
    \STATE $\bz^{(t+1)} = \bz^{(t)} + \gamma \bs\left(\bz^{(t)}\right) + \tfrac{\gamma}{\rho^2} \left(\bx^{(t)} - \bz^{(t)}\right) + \sqrt{2\gamma} \boldsymbol{\varepsilon}^{(t)}$
    \algocomment{Sampling the variable of interest according to \eqref{gaussian_step}}
    \STATE  $\bx^{(t+1)} \sim \mathcal{N}\big( \boldsymbol{\mu}(\bz^{(t+1)}), \mathbf{Q}^{-1} \big)$
\ENDFOR
\renewcommand{\algorithmicensure}{\textbf{Output:}}
\ENSURE collection of samples $\{\bx^{(t)}, \bz^{(t)}\}_{t=N_{\mathrm{bi}+1}}^{N_{\mathrm{MC}}}$
\end{algorithmic}
\end{algorithm}

\subsection{Score function-based priors}\label{subsec:priors}
The LwSGS algorithm reviewed above appears to be particularly well suited to sample from posterior distributions once their associated log-priors are differentiable with explicit or easy-to-compute gradients. Among the priors satisfying this property, noticeable prototypes can be easily exhibited from the literature. Taking their roots in various communities with distinct motivations, they all share the property of being defined by a score function, either implicitly or explicitly. Four of them are discussed below in the context of Bayesian IR.

\subsubsection{Regularization by denoising (RED) priors}
Romano \emph{et al.} introduced the RED framework to exploit denoisers as explicit regularizations \cite{romano2017little}. The RED engine relies on a denoiser $\mathsf{D}: \mathbb{R}^n \rightarrow \mathbb{R}^n$ and defines the regularization as
\begin{equation}\label{RED}
       g_{\textrm{red}}(\bx) = \tfrac{\beta}{2}\bx^T  \left( \bx - \mathsf{D}(\bx) \right)
\end{equation}
where $\beta$ adjusts the amount of regularization. RED offers great flexibility in choosing the denoiser, in particular leveraging pre-trained deep NNs. However, it requires that $\mathsf{D}(\cdot)$ is differentiable, locally homogeneous, Jacobian symmetric, and satisfies the passivity condition \cite{romano2017little,reehorst2018regularization}. Traditionally applied in a deterministic optimization context, RED's ability to define complex priors without explicitly defining them has made it a popular choice for various image restoration (IR) tasks. In a probabilistic setting, RED can be encapsulated into a Bayesian model by defining an appropriate prior distribution from the RED potential
\begin{equation}\label{RED_prior}
     p_{\textrm{red}}(\bx) \propto \exp \left[ -\tfrac{\beta}{2}\bx^T  \left( \bx - \mathsf{D}(\bx) \right) \right].    
\end{equation}
Formally, let the matrix $\boldsymbol{\Lambda}(\bx) = \bI_n - \nabla \mathsf{D}(\bx)$, for all $\bx \in \mathbb{R}^n$. If $\boldsymbol{\Lambda}(\bx)$ has at least one non-zero eigenvalue and the denoiser $\mathsf{D}(\cdot)$ satisfies the differentiability, local homogeneity, Jacobian symmetry and passivity conditions, then the function $p_{\textrm{red}}(\cdot)$ defines a proper probability density function (pdf). Moreover, the score function associated with $p_{\textrm{red}}(\cdot)$ is expressed as
\begin{equation}\label{RED_score}
{\bs}_{\textrm{red}}(\bx)  = - \beta \left(\bx - \mathsf{D}(\bx)\right).
\end{equation}
In other words, the  RED potential specified by \eqref{RED} leads to a score function defined as a denoising residual. At the first look, this identity seems to be only intrinsic to the RED paradigm, motivated by some sound yet empirical considerations related to appealing properties of any denoiser \cite{milanfar2024denoising}. However, the identity \eqref{RED_score} deserves to be envisioned from a much broader perspective to realize that one of its particular instances is a keystone to relate various prior models. Indeed, the Tweedie's formula states that this identity is always satisfied when considering the score function $\bs_{\eta}(\bx)= \grad \log\, p_{\eta}(\bx) $ associated with a smooth counterpart $p_{\eta}(\bx) \propto \int p(\tilde{\bx}) \exp\left(-\tfrac{1}{2\eta ^2}\left\|\bx-\tilde{\bx}\right\|^2\right) \mathsf{d}\tilde{\bx}$ of the prior once the operator $\mathsf{D}(\cdot)$ is chosen as a MMSE denoiser. Conceptually, this means that learning an MMSE denoiser is equivalent of learning a given score function. This formula also allows existing deterministic PnP algorithms to be re-derived while embedding  score functions in place of  conventional deterministic denoisers, offering the ability of resorting to score-based models \cite{park2024plug,Coeurdoux2024pnp}. These generative models are also considered in the framework investigated herein and described in the next section.


\subsubsection{Score-based generative model (SGM) priors}
SGMs aim to approximate a target distribution $\pi(\bx)$ by an instrumental distribution $p_{\textrm{sgm}}(\bx;\boldsymbol{\theta})$ using a score matching strategy \cite{song2019generative}. As stated in Section \ref{sec:score}, it consists in minimizing the Fisher divergences between the target and instrumental densities according to \eqref{eq:score_matching} \cite{hyvarinen2005estimation}. In practical scenarios, solving this optimization problem   faces a major problem: in general, the (gradient of the log-) target distribution is not known explicitly but is only available through a set of samples assumed to be drawn from $\pi(\bx)$.
However, as suggested by the equivalence highlighted in the previous paragraph, training an MMSE denoiser and a score function is expected to be feasible in similar experimental settings. Indeed, both approaches can leverage pairs of noisy and noise-free images. More precisely, a noise conditional score SGM $\bs_{\textrm{sgm}}(\bx;\boldsymbol{\theta})$ can be trained from a set of available images $\bx$ drawn from $\pi(\bx)$ and a set of images $\tilde{\bx}$ corrupted by additive white Gaussian noise, i.e., $\tilde{\bx} = \bx + \eta \bw$ with $\bw \sim \mathcal{N}\left(\boldsymbol{0}, \bI\right)$, by solving \cite{vincent2011connection}
\begin{equation}\label{eq:SGM_training}
\operatornamewithlimits{min}_{\boldsymbol{\theta}} \mathbb{E}_{\bx,\tilde{\bx}}\left[ \left\| \tfrac{1}{\eta^2}\left(\tilde{\bx} - \bx\right) + {\bs}_{\textrm{sgm}}(\tilde{\bx};\boldsymbol{\theta})  \right\|^2  \right]
\end{equation}
where $\boldsymbol{\theta}$ represents the NN parameters. In practice, better generation abilities of SGM are achieved by learning a set $ {\bs}_{\textrm{sgm}}(\cdot,j;\boldsymbol{\theta})$  ($j=1,\ldots,J$)  of conditional score NNs defined over a range of $J$ scales of noise perturbations $\eta_1^2 < \ldots < \eta_J^2$. In this case the training loss function \eqref{eq:SGM_training} is replaced by a weighted sum of Fisher divergences rewritten as
\begin{equation}\label{eq:SGM_training_average}
\operatornamewithlimits{min}_{\boldsymbol{\theta}} \sum_{j=1}^J \alpha_j \mathbb{E}_{\bx,\tilde{\bx}_j}\left[ \left\| \tfrac{1}{\eta_j^2}\left(\tilde{\bx}_j - \bx\right) + {\bs}_{\textrm{sgm}}(\tilde{\bx}_j,j;\boldsymbol{\theta})  \right\|^2  \right]
\end{equation}
where $\tilde{\bx}_j=\tilde{\bx}+\eta_j \bw$ and $\alpha_j$ are weighting factors that can be commonly chosen as $\alpha_j=\eta^2_j$. Interestingly this loss function is precisely the one derived to train a particular instance of diffusion-based models, namely the denoising diffusion probabilistic model (DDPM) \cite{ho2020denoising}. Although these latter generative models emerge from significantly different rationales, i.e., exploiting a stochastic differential equation (SDE) to progressively denoise images corrupted by infinitesimal white noise, DDPM has  been shown to be equivalent to SGM \cite{song2021score}. Once trained, an SGM can be used to draw new samples from the instrumental distribution $p_{\textrm{sgm}}(\bx)$ using ULA \cite{song2021score}. This simple scheme can be easily adapted to sample from the posterior distribution associated with a Bayesian inverse problem, with a prior score function defined as ${\bs}_{\textrm{sgm}}({\bx})$.

\subsubsection{Normalizing flow (NF) priors}
Normalizing flows define a class of differentiable and invertible NNs that encode a mapping $T: \mathbb{R}^n \rightarrow \mathbb{R}^n$ from a known (e.g., Gaussian) latent distribution $q({\bu})$ towards an instrumental distribution $p_{\textrm{nf}}(\bx)$ approximating a possibly more complex target distribution $\pi(\bx)$ \cite{kobyzev2020normalizing, kingma2018glow}. Once the NN has been trained using samples drawn from $\pi(\bx)$, the instrumental distribution of $\bx = T({\bu})$ can be explicitly derived following the change of variable formula
\begin{equation*}
	\log\,p_{\textrm{nf}}(\bx) = \log\, q\left( T^{-1}(\bx) \right) +  \log \,|J_{T^{-1}} (\bx) |
\end{equation*}
where $J_{T} (\cdot)$ is the $n \times n$ Jacobian matrix of the mapping $T(\cdot)$. The key benefit of using NF-based generative models lies in the closed-form expression of the instrumental distribution, although this comes at the cost of evaluating the inverse of the learned mapping and its determinant, which can be computationally expensive. However, the particular architectures of NF proposed in the literature allow these computations to be lightened, e.g., thanks to the use of affine coupling layers. Once the NF has been trained, evaluating the score function $\bs_{\textrm{nf}}({\bx})$ associated with $p_{\textrm{nf}}(\bx)$ can be easily achieved using automatic differentiation libraries of deep learning frameworks. It is worth noting that the neural ordinary differential equation associated with the diffusion-based SDE discussed in the previous paragraph defines a continuous instance of NF \cite{chen2018neural,grathwohl2019scalable}. Finally, sampling from the resulting prior distribution can be easily achieved by ULA in the target (image) domain \cite{cai2023nf} or by more advanced techniques in the latent domain \cite{Coeurdoux_MLJ_2024}.

\subsubsection{Convex-ridge regularizers (CRR) priors}
Despite their popularity, standard convex regularizers are often outperformed by data-driven regularizations, particularly those based on convolutional NNs (CNNs). Recently, CNN-based CRRs have been proposed as 
\begin{equation*}
    \log \, p_{\textrm{crr}}(\bx) = - g_{\textrm{crr}}(\bx) = - \sum_{i=1}^{n\times N_C} \phi_i({\bw}_i^\top \bx)
\end{equation*}
where $\phi_i: \mathbb{R} \rightarrow \mathbb{R}$ are convex functions and $\textbf{w}_i \in \mathbb{R}^n$ are learnable weights \cite{goujon2023neural}. The corresponding score function is expressed as a bias-free 1-layer CNN
\begin{equation}\label{eq:CRR}
   \bs_{\textrm{crr}}(\bx) = -{\bW}^\top \mathsf{act}({\bW} \bx)
\end{equation}
where ${\bW}=[{\bw}_1 \cdots {\bw}_{nN_C}] \in \mathbb{R}^{n N_C \times n}$ and $N_C$ is the number of image channels. In \eqref{eq:CRR}, $\mathsf{act}(\cdot) = \left[\phi^{\prime}_i(\cdot)\right]_{i=1}^{nN_C}$ denotes an activation function that can be parametrized with trainable non-decreasing and non-expansive linear splines.
Despite the apparent unrelatedness of CRR with the previously discussed score function-based prior models, some important connections can be drawn. In \cite{goujon2023neural}, the gradient of the CRR is classically used to define a gradient-step operator which is trained on pairs of noise-free and noisy images. Alternatively, by interpreting this gradient as a noise estimator, another possible  training strategy would consist in considering the SGM minimization problem \eqref{eq:SGM_training}. Moreover, the score in \eqref{eq:CRR} also corresponds to the output of an autoencoder (AE) defined by a bias-free single hidden layer. Adopting this alternate interpretation draws a direct connection to SGM discussed above. Indeed, up to a constant term, the score matching strategy employed to train an SGM is shown to be equivalent to  training a denoising AE \cite{vincent2008extracting,vincent2010stacked}, i.e., an AE which aims at recovering noise-free versions of  noisy input images \cite{vincent2011connection}. Finally, it is worth noting that  CRR-based priors has been already embedded into a sampling scheme targeting a posterior distribution of the form \eqref{posterior} to perform Bayesian image deconvolution \cite{klatzer2023accelerated}.


\section{Experiments on simulated data sets}\label{sec:experiments}
\subsection{Experimental setup}
Image inpainting and single image super-resolution are two tasks considered to illustrate the effectiveness of the LwSGS algorithm when embedding the score function-based priors considered in Section \ref{subsec:priors} . These experiments are performed on a set of $30$ RGB images of size $256 \times 256$ pixels ($n=256^2$) from the Flickr Faces High Quality (FFHQ) data set \cite{karras2019style} and the ImageNet data set \cite{deng2009imagenet}.  All images have been normalized to the range $[0,1]$. For the inpainting task, the operator $\bA \in \left\{0,1\right\}^{m \times n}$ ($m \ll n$) denotes a binary mask where $80\%$ of pixels are randomly masked across all color channels. Conversely,  for the single image super-resolution task, the operator $\bA$ is decomposed as $\bA = \bS \bB$. Here, the ${n \times n}$ matrix $\bB$ represents a spatially invariant Gaussian blur of size $7 \times 7$ with standard deviation  $1.6$, while  the operator $\bS  \in \left\{0,1\right\}^{m \times n}$ is a binary matrix performing a regular subsampling with a factor $d=4$ in each direction (i.e., $n = m d^2$). For the two considered tasks, the degraded images have been corrupted by additive Gaussian noise to achieve a signal-to-noise ratio ($\mathrm{SNR}$) of $30$dB. 

\subsection{Compared methods}
The instances of the LwSGS algorithm are denoted as $\mathcal{P}$-LwSGS where $\mathcal{P} \in \left\{\text{RED},\text{NF},\text{SGM},\text{CRR} \right\}$ refers to the considered score-based prior. They are compared to existing counterparts, namely PnP-ULA \cite{laumont2022bayesian} and NF-ULA \cite{cai2023nf}, as well as newly designed counterparts SGM-ULA and CRR-ULA which sample according to the corresponding posterior distributions with ULA. Pre-trained models without further fine-tuning are used to define the prior distributions: DRUNet\footnote{\url{https://github.com/cszn/DPIR}} for the PnP/RED prior \cite{zhang2021plug}, patch normalizing flow\footnote{\url{https://github.com/FabianAltekrueger/patchNR}} (patchNR) for the NF-based prior \cite{altekruger2023patchnr}, the SGM-based model\footnote{\url{https://github.com/yuanzhi-zhu/DiffPIR}} proposed by \cite{zhu2023denoising} and the CNN-based CRR\footnote{\url{https://github.com/axgoujon/convex_ridge_regularizers}} proposed by \cite{goujon2023neural}.
The results reported below correspond to the minimum mean square error (MMSE) estimates approximated by averaging the generated samples produced by the considered sampling schemes, i.e., 
\begin{equation}\label{eq:MMSE}
\hat{\mathbf{x}}_{\mathrm{MMSE}} = \frac{1}{N_{\mathrm{MC}}-N_{\mathrm{bi}}} \sum_{t=N_{\mathrm{bi}}+1}^{N_{\mathrm{MC}}} \mathbf{x}^{(t)}
\end{equation}
 where $N_{\mathrm{MC}}$  is the total number of iterations including $N_{\mathrm{bi}}$ burn-in iterations. For the inpainting task, the coupling parameter has been fixed as $\rho^2 = 1$ for $\left\{\text{RED},\text{SGM},\text{NF}\right\}$-LwSGS and $\rho^2=10^{-5}$ for CRR-LwSGS. For the super-resolution task, the LwSGS method should be slightly adapted to follow a double splitting \cite{Faye2024}. For the two tasks, the total number of iterations has been set as $N_{\mathrm{MC}}=7\,500$ for $\left\{\text{RED},\text{SGM}\right\}$-LwSGS and PnP-ULA and $N_{\mathrm{MC}}=25\,000$ for $\left\{\text{CRR},\text{NF} \right\}$-$\left\{\text{LwSGS},\text{ULA}\right\}$. This difference can be explained by the fact that PatchNR and CRR are less expressive priors requiring more iterations to reach satisfying performances.  

\begin{table*}
\setlength{\tabcolsep}{5pt}
\centering
\caption{FFHQ and ImageNet data sets: average performance and corresponding standard deviations.}
\label{Tab:perfs_ffhq_imagenet}
\resizebox{\columnwidth}{!}{
\begin{tabular}{lllcccccccc} 
\toprule
& & & \multicolumn{2}{c}{{\textbf{RED}}} & \multicolumn{2}{c}{{\textbf{NF}}} & \multicolumn{2}{c}{{\textbf{SGM}}} &\multicolumn{2}{c}{{\textbf{CRR}}} \\
\toprule
& & & {\textbf{LwSGS}} & {\textbf{ULA}} & {\textbf{LwSGS}} & {\textbf{ULA}} & {\textbf{LwSGS}} & {\textbf{ULA}} & {\textbf{LwSGS}} & {\textbf{ULA}} \\
\midrule
\multirow{10}{*}{\rotatebox{90}{{\textbf{FFHQ}}}} & \multirow{5}{*}{\rotatebox{90}{{\textbf{Inpainting}}}} &
 {\textbf{PSNR}}(dB)$\uparrow$ & 30.88±3.04 & 31.17±3.26 & 30.28±1.69 & 30.17±1.67 & 30.30±2.25 & 29.37±1.76 & 29.45±1.86 & 29.37±1.85\\
& & {\textbf{SSIM}}$\uparrow$ & 0.915±0.020 & 0.908±0.022 & 0.869±0.018 & 0.867±0.019 & 0.901±0.016 & 0.879±0.020 & 0.876±0.021 & 0.876±0.022 \\
& & {\textbf{LPIPS}} $\downarrow$ & 0.127±0.026 & 0.143±0.022 & 0.194±0.027 & 0.191±0.028 & 0.103±0.019 & 0.161±0.023 & 0.184±0.022 & 0.186±0.023 \\
& & {\textbf{IAT}} $\downarrow$ & 74.59±11.08 & 74.71±11.16 & 75.86±8.50 & 75.81±8.51 & 74.89±8.30 & 76.61±8.50 & 76.26±11.02 & 76.16±11.02   \\
& & {\textbf{Time}}(s) $\downarrow$ & 55±1 & 57±0 & 1730±64 & 1757±139 & 156±1  & 175±1 & 51±1 & 53±1 \\
\cmidrule{2-11}
& \multirow{5}{*}{\rotatebox{90}{{\textbf{Super-res.}}}} &
 {\textbf{PSNR}}(dB)$\uparrow$ & 30.75±2.05 & 29.98±1.99 & 29.46±2.02 & 28.97±2.10 & 29.28±1.96 &  28.36±1.99 & 29.60±2.00 & 29.56±1.99  \\
& & {\textbf{SSIM}}$\uparrow$ & 0.880±0.026 & 0.869±0.027 & 0.843±0.026 & 0.842±0.030 & 0.854±0.029 & 0.832±0.034 & 0.858±0.030 & 0.857±0.029 \\
& & {\textbf{LPIPS}} $\downarrow$ & 0.168±0.027 & 0.213±0.031 & 0.241±0.025 & 0.263±0.029 & 0.238±0.029 & 0.267±0.034 & 0.221±0.032 & 0.223±0.031 \\
& & {\textbf{IAT}} $\downarrow$ & 76.41±8.39 & 77.79±8.50 & 77.02±10.80 & 78.21±11.10 & 77.86±8.55 & 80.31±8.71 & 77.30±8.44 & 77.52±8.47   \\
& & {\textbf{Time}}(s) $\downarrow$ & 54±4 & 66±8 & 1351±803 & 1378±791 & 473±0 & 487±4 & 61±1 & 65±3 \\
\midrule
\midrule
\multirow{10}{*}{\rotatebox{90}{{\textbf{ImageNet}}}} & \multirow{5}{*}{\rotatebox{90}{{\textbf{Inpainting}}}} &
 {\textbf{PSNR}}(dB)$\uparrow$ & 30.12±3.75 & 30.32±3.27 & 29.88±2.96 & 29.68±2.93 & 31.27±3.47 & 30.72±3.37 & 28.96±2.92 & 28.91±2.99 \\
& & {\textbf{SSIM}}$\uparrow$ & 0.872±0.111 & 0.875±0.083 & 0.846±0.073 & 0.836±0.078 & 0.884±0.078 & 0.864±0.087 & 0.846±0.081 & 0.845±0.083 \\
& & {\textbf{LPIPS}} $\downarrow$ & 0.149±0.072 & 0.202±0.077 & 0.217±0.060 & 0.221±0.058 & 0.151±0.063 & 0.217±0.072 & 0.221±0.052 & 0.222±0.052 \\
& & {\textbf{IAT}} $\downarrow$ & 75.62±10.99 & 76.46±12.45 & 77.61±12.53 & 77.68±12.54 & 75.63±12.48 & 75.87±11.99 & 76.58±12.33 & 76.60±12.32  \\
& & {\textbf{Time}}(s) $\downarrow$ & 154±8 & 179±14 & 1548±441 & 1592±536 & 1009±19  & 1196±28 & 52±0 & 56±3 \\
\cmidrule{2-11}
& \multirow{5}{*}{\rotatebox{90}{{\textbf{Super-res.}}}} &
 {\textbf{PSNR}}(dB)$\uparrow$ & 29.57±3.94 & 28.80±3.55 & 28.52±3.43 & 28.47±3.46 & 28.39±3.44 &  28.24±3.35 & 28.71±3.60 & 28.59±3.52  \\
& & {\textbf{SSIM}}$\uparrow$ & 0.826±0.107 & 0.817±0.107 & 0.803±0.106 & 0.793±0.103 & 0.807±0.108 & 0.800±0.106 & 0.813±0.108 & 0.811±0.107 \\
& & {\textbf{LPIPS}} $\downarrow$ & 0.218±0.090 & 0.254±0.089 & 0.279±0.072 & 0.278±0.077 & 0.268±0.081 & 0.271±0.082 & 0.249±0.087 & 0.251±0.086\\
& & {\textbf{IAT}} $\downarrow$ & 75.35±10.98 & 76.07±11.04 & 76.51±11.30 & 76.91±10.24 & 76.11±11.04 & 76.93±10.13 & 75.77±11.14 & 75.99±11.30   \\
& & {\textbf{Time}}(s) $\downarrow$ & 51±1 & 57±0 & 1805±289 & 2006±280 & 1189±290 & 1224±293 & 57±1 & 60±1 \\
\bottomrule
\end{tabular}
}
\end{table*}

\subsection{Figures-of-merit}
To quantitatively compare the performance of the algorithms, several widely used metrics are considered, namely the peak signal-to-noise ratio (PSNR) (dB), the structural similarity index (SSIM) \cite{wang2004image}, as well as the learned perceptual image patch similarity (LPIPS) \cite{zhang2018unreasonable}. Specifically, PSNR is a global image metric which measures the faithfulness of restoration between two images. Furthermore, SSIM is used as a spatial metric  measuring the structural similarity between two given images. LPIPS is a perceptual metric that quantifies the human-perceived similarity between two images. Moreover, the effectiveness of the sampling algorithms is assessed by comparing the integrated autocorrelation time (IAT), with lower values signifying better mixing performance \cite{fall2025simple}. Finally, the algorithms are compared in terms of computational times when run on a server equipped with 48 Intel 2.8GHz CPU cores, 384GB RAM, and Nvidia A100 GPU.

\subsection{Experimental results}\label{subsec:results}
Table \ref{Tab:perfs_ffhq_imagenet} reports the average PSNR (dB), SSIM, LPIPS and IAT results of the compared methods when performing the two considered tasks on the FFHQ and Imagenet data sets. These results demonstrate that the LwSGS algorithm achieves highly competitive performances. It produces high-quality reconstructions that closely match ground-truth details. Specifically, for the inpainting task, the proposed framework effectively fills in missing regions with remarkable accuracy, as demonstrated in Fig. \ref{fig:inpainting}. Similarly, for the super-resolution task, LwSGS enhances image resolution while preserving intricate details, as shown in Figure \ref{fig:superresolution}. When compared to ULA, LwSGS is shown to provide consistently better results. Specifically, when employing SGM and RED as prior models, images reconstructed using the LwSGS-based sampling method exhibit closer fidelity to the ground truth, showcasing sharper details. Even when utilizing NF- and CRR-based priors, the LwSGS framework consistently outperforms ULA, albeit with a narrower margin in performance difference. LwSGS consistently achieves competitive performance across most metrics, often outperforming or closely matching ULA with the various priors.

\newcommand{\subplotwidth}{.145\textwidth}
\newcommand{\subplotwidthbis}{.1\textwidth}
\newcommand{\subplotwidthbisbis}{.04\textwidth}
\begin{figure*}
\centering
\begin{tabularx}{0.9\textwidth} { 
   >{\centering\arraybackslash}X 
   >{\centering\arraybackslash}X 
   >{\centering\arraybackslash}X 
   >{\centering\arraybackslash}X 
   >{\centering\arraybackslash}X 
   >{\centering\arraybackslash}X 
   }
    \scriptsize{Ground Truth} &  \scriptsize{Observation} & \scriptsize{RED-LwSGS}  & \scriptsize{SGM-LwSGS} &  \scriptsize{NF-LwSGS} & \scriptsize{CRR-LwSGS} 
\end{tabularx}\\%
\begin{tikzpicture}[zoomboxarray]
    \node [image node] { \includegraphics[width=\subplotwidth]{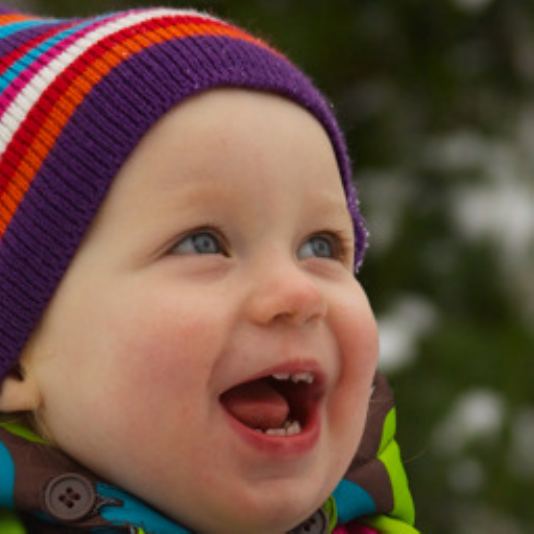}};
    \zoombox{0.6, 0.54}
\end{tikzpicture}
\begin{tikzpicture}[zoomboxarray]
    \node [image node] { \includegraphics[width=\subplotwidth]{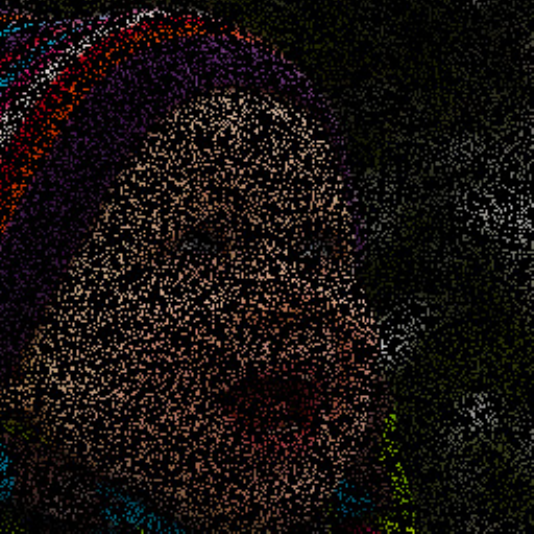}};
    \zoombox{0.6, 0.54}
\end{tikzpicture}
\begin{tikzpicture}[zoomboxarray]
    \node [image node] { \includegraphics[width=\subplotwidth]{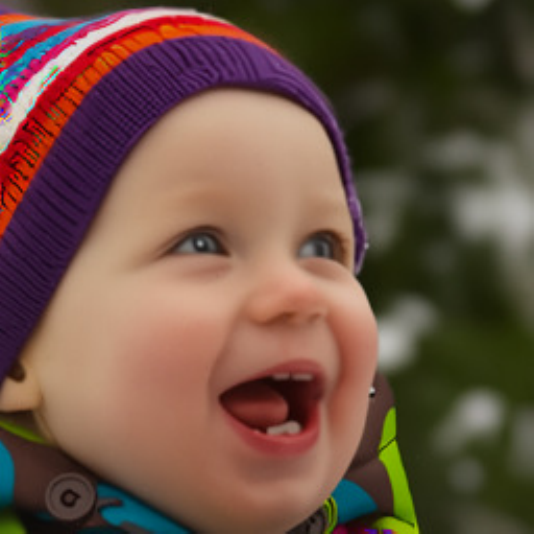}};
    \zoombox{0.6, 0.54}
\end{tikzpicture}
\begin{tikzpicture}[zoomboxarray]
    \node [image node] { \includegraphics[width=\subplotwidth]{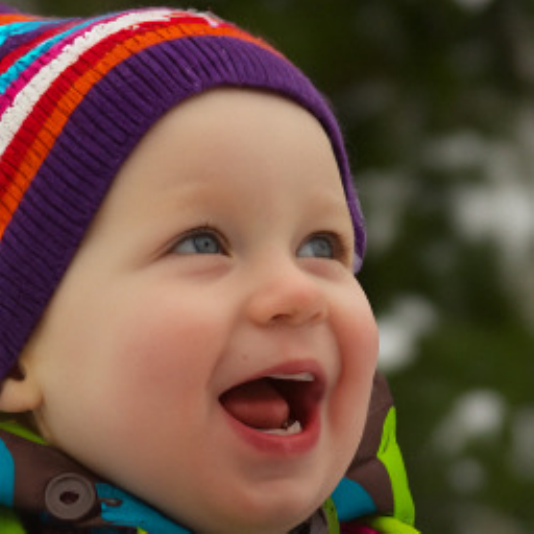}};
    \zoombox{0.6, 0.54}
\end{tikzpicture}
\begin{tikzpicture}[zoomboxarray]
    \node [image node] { \includegraphics[width=\subplotwidth]{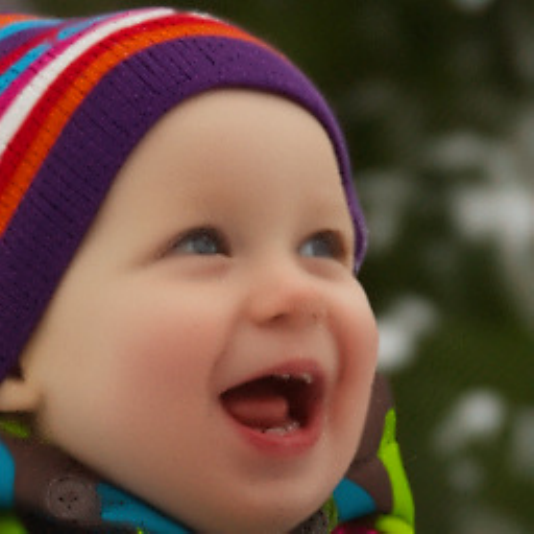}};
    \zoombox{0.6, 0.54}
\end{tikzpicture}
\begin{tikzpicture}[zoomboxarray]
    \node [image node] { \includegraphics[width=\subplotwidth]{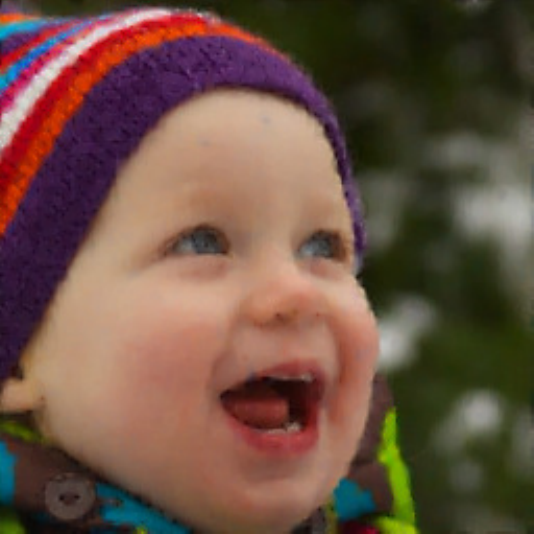}};
    \zoombox{0.6, 0.54}
\end{tikzpicture}\\
\begin{tikzpicture}[zoomboxarray]
    \node [image node] { \includegraphics[width=\subplotwidth]{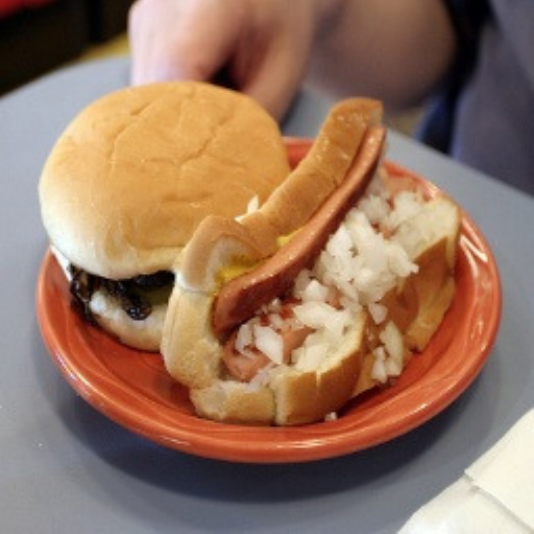}};
    \zoombox{0.6, 0.6}
\end{tikzpicture}
\begin{tikzpicture}[zoomboxarray]
    \node [image node] { \includegraphics[width=\subplotwidth]{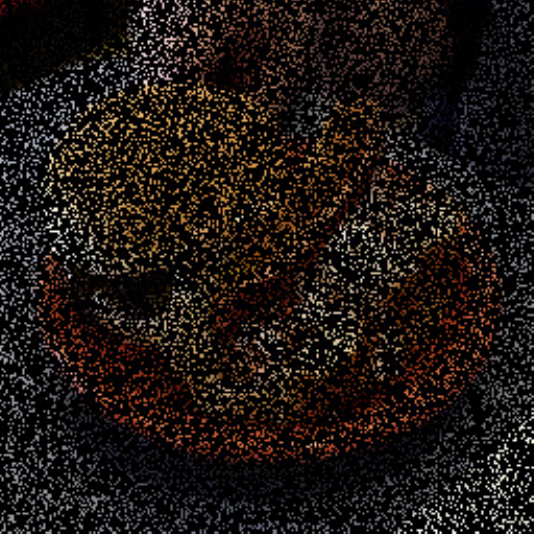}};
    \zoombox{0.6, 0.6}
\end{tikzpicture}
\begin{tikzpicture}[zoomboxarray]
    \node [image node] { \includegraphics[width=\subplotwidth]{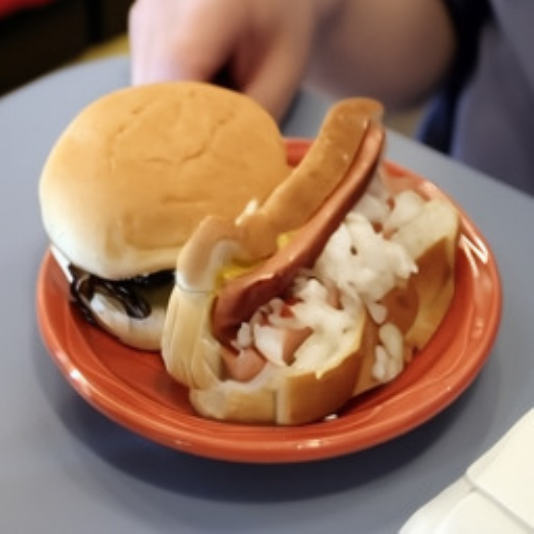}};
    \zoombox{0.6, 0.6}
\end{tikzpicture}
\begin{tikzpicture}[zoomboxarray]
    \node [image node] { \includegraphics[width=\subplotwidth]{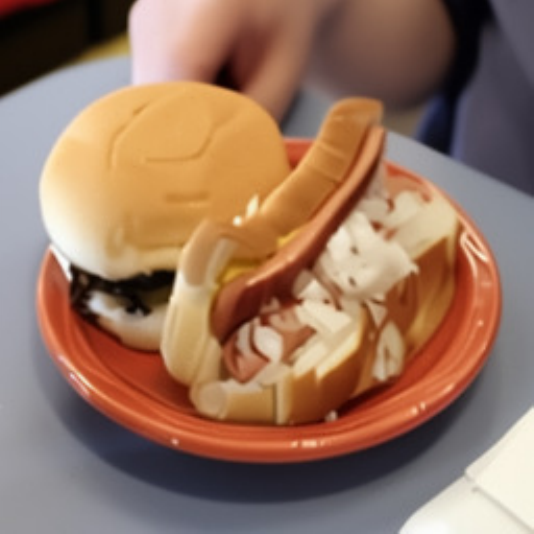}};
    \zoombox{0.6, 0.6}
\end{tikzpicture}
\begin{tikzpicture}[zoomboxarray]
    \node [image node] { \includegraphics[width=\subplotwidth]{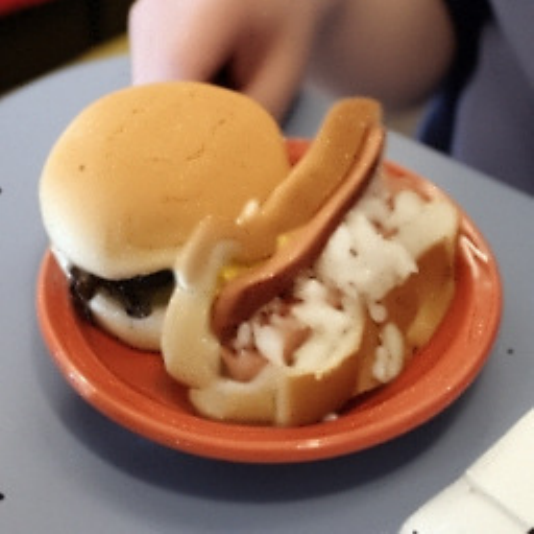}};
    \zoombox{0.6, 0.6}
\end{tikzpicture}
\begin{tikzpicture}[zoomboxarray]
    \node [image node] { \includegraphics[width=\subplotwidth]{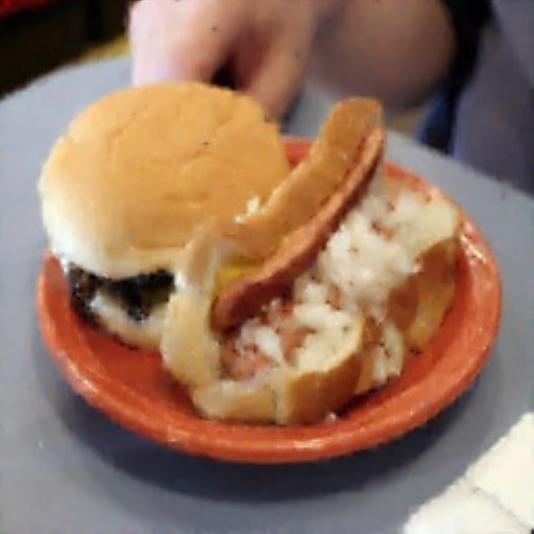} };
    \zoombox{0.6, 0.6}
\end{tikzpicture}%
\caption{Inpainting task on the FFHQ (top) and ImageNet data sets (bottom).}
\label{fig:inpainting}
\end{figure*}

\begin{figure*}
\centering
\begin{tabularx}{0.9\textwidth} { 
   >{\centering\arraybackslash}X 
   >{\centering\arraybackslash}X 
   >{\centering\arraybackslash}X 
   >{\centering\arraybackslash}X 
   >{\centering\arraybackslash}X 
   >{\centering\arraybackslash}X 
   }
    \scriptsize{Ground Truth} &  \scriptsize{Observation} & \scriptsize{RED-LwSGS}  & \scriptsize{SGM-LwSGS} &  \scriptsize{NF-LwSGS} & \scriptsize{CRR-LwSGS} 
\end{tabularx}

\begin{tikzpicture}[zoomboxarray]
    \node [image node] { \includegraphics[width=\subplotwidth]{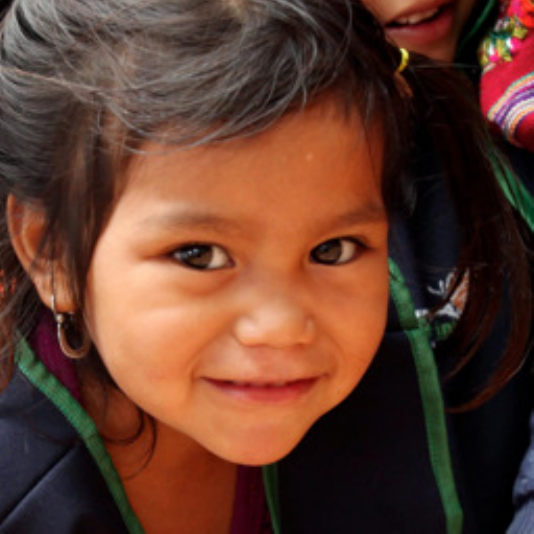}};
    \zoombox{0.64, 0.54}
\end{tikzpicture}
\begin{tikzpicture}[zoomboxarray]
    \node [image node] { \includegraphics[width=\subplotwidth]{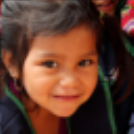}};
    \zoombox{0.64, 0.54}
\end{tikzpicture}
\begin{tikzpicture}[zoomboxarray]
    \node [image node] { \includegraphics[width=\subplotwidth]{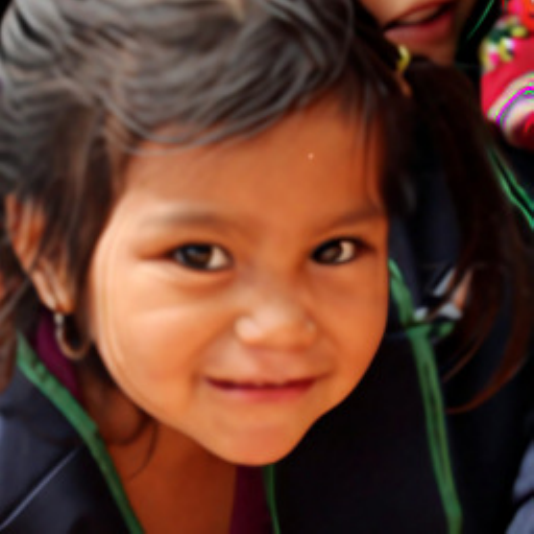}};
    \zoombox{0.64, 0.54}
\end{tikzpicture}
\begin{tikzpicture}[zoomboxarray]
    \node [image node] { \includegraphics[width=\subplotwidth]{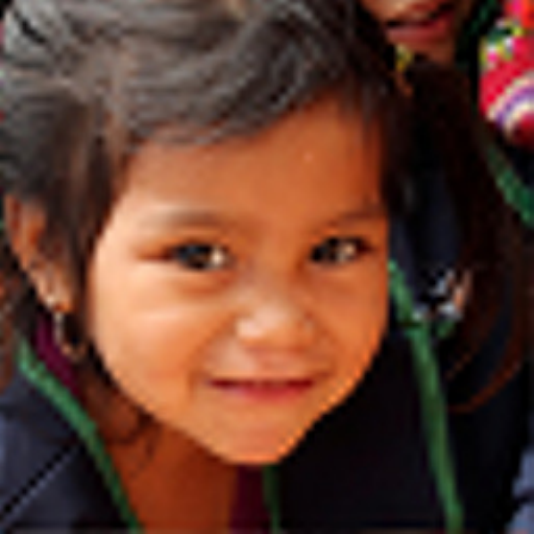}};
    \zoombox{0.64, 0.54}
\end{tikzpicture}
\begin{tikzpicture}[zoomboxarray]
    \node [image node] { \includegraphics[width=\subplotwidth]{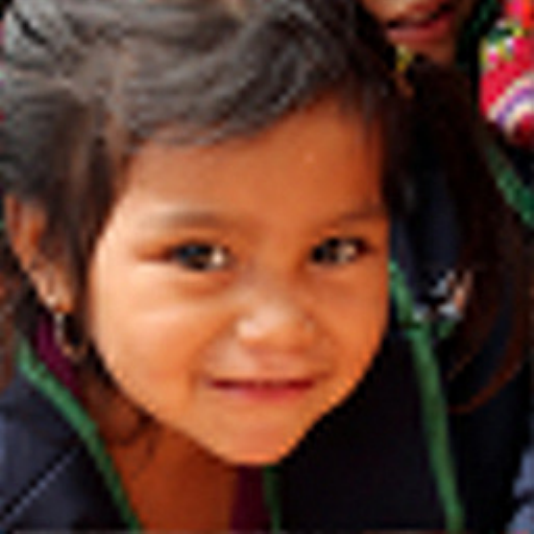}};
    \zoombox{0.64, 0.54}
\end{tikzpicture}
\begin{tikzpicture}[zoomboxarray]
    \node [image node] { \includegraphics[width=\subplotwidth]{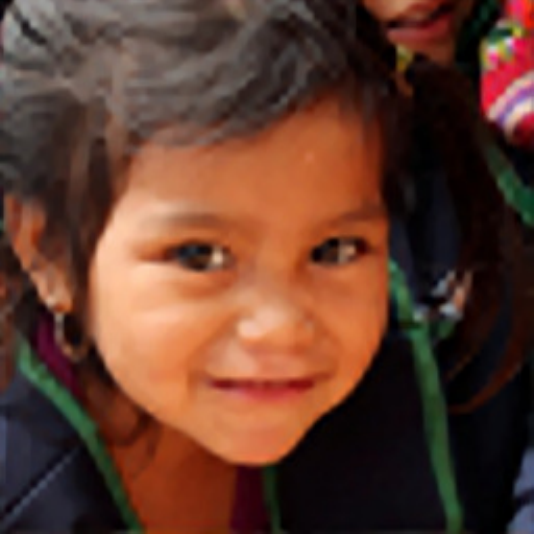}};
    \zoombox{0.64, 0.54}
\end{tikzpicture}
      \\
\begin{tikzpicture}[zoomboxarray]
    \node [image node] {\includegraphics[width=\subplotwidth]{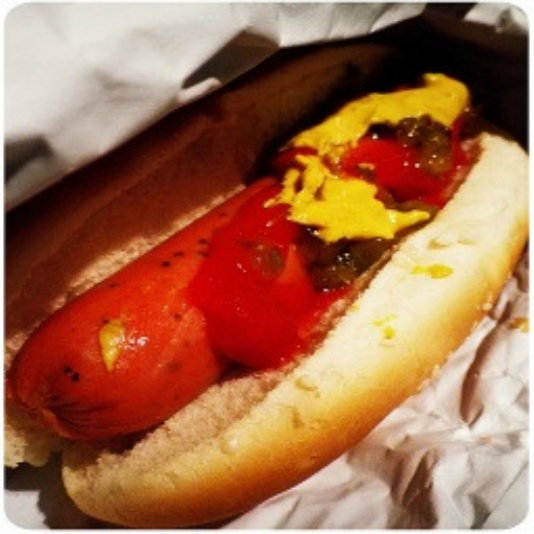}};
    \zoombox{0.64, 0.54}
\end{tikzpicture} 
\begin{tikzpicture}[zoomboxarray]
    \node [image node] {\includegraphics[width=\subplotwidth]{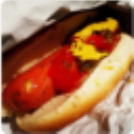}};
    \zoombox{0.64, 0.54}
\end{tikzpicture} 
\begin{tikzpicture}[zoomboxarray]
    \node [image node] {\includegraphics[width=\subplotwidth]{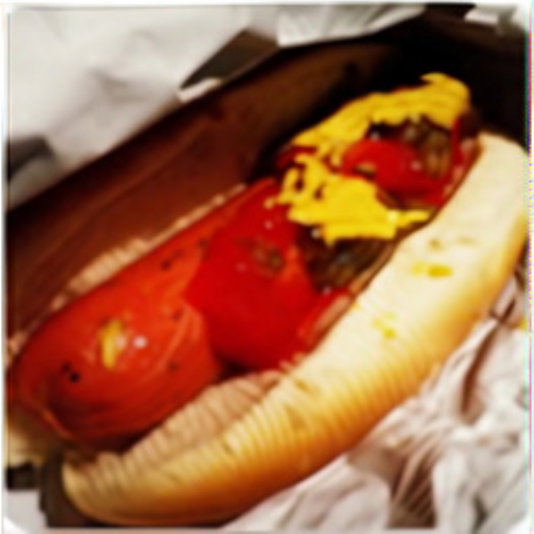}};
    \zoombox{0.64, 0.54}
\end{tikzpicture} 
\begin{tikzpicture}[zoomboxarray]
    \node [image node] {\includegraphics[width=\subplotwidth]{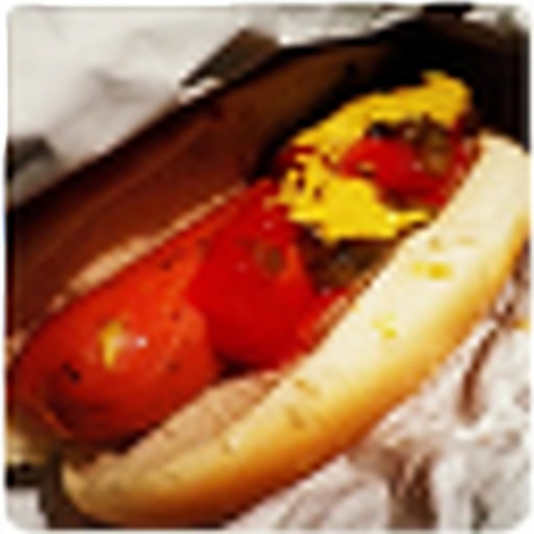}};
    \zoombox{0.64, 0.54}
\end{tikzpicture} 
\begin{tikzpicture}[zoomboxarray]
    \node [image node] {\includegraphics[width=\subplotwidth]{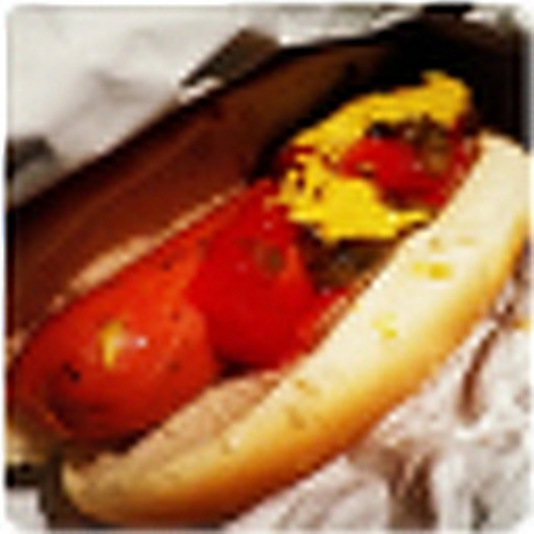}};
    \zoombox{0.64, 0.54}
\end{tikzpicture} 
\begin{tikzpicture}[zoomboxarray]
    \node [image node] {\includegraphics[width=\subplotwidth]{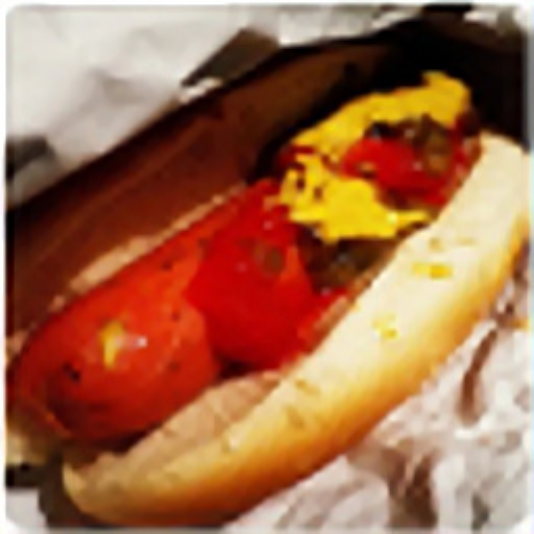}};
    \zoombox{0.64, 0.54}
\end{tikzpicture}%
\caption{Super resolution task on the FFHQ (top) and ImageNet data sets (bottom).}
\label{fig:superresolution}
\end{figure*}

Additionally, as previously mentioned, LwSGS generates samples that asymptotically follow the posterior distribution, enabling the quantification of uncertainty inherent to the obtained point estimates. This capability is illustrated in Fig. \ref{fig:imagenet} where estimated pixelwise standard deviations are depicted for the LwSGS-based algorithms when performing the inpainting tasks. The uncertainty is notably higher for pixels that were inpainted, which aligns with expectations for inpainting tasks.

\renewcommand{\subplotwidth}{.145\textwidth}
\newcommand{\subplotwidthbar}{.115\textwidth}
\renewcommand{\subplotwidthbis}{.1\textwidth}
\renewcommand{\subplotwidthbisbis}{.04\textwidth}
\begin{figure*}
\centering
\begin{tabularx}{0.9\textwidth} { 
   >{\centering\arraybackslash}X 
   >{\centering\arraybackslash}X 
   >{\centering\arraybackslash}X 
   >{\centering\arraybackslash}X 
   >{\centering\arraybackslash}X 
   >{\centering\arraybackslash}X 
   }
    \scriptsize{Ground Truth} &  \scriptsize{Observation} & \scriptsize{RED-LwSGS}  & \scriptsize{SGM-LwSGS} &  \scriptsize{NF-LwSGS} & \scriptsize{CRR-LwSGS} 
\end{tabularx}\\%
\begin{tikzpicture}[zoomboxarray]
    \node [image node] {\includegraphics[width=\subplotwidth]{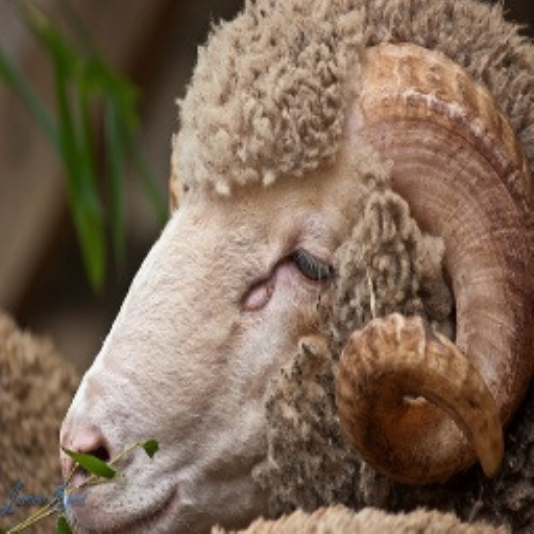}};
    \zoombox{0.58, 0.50}
\end{tikzpicture}
\begin{tikzpicture}[zoomboxarray]
    \node [image node] {\includegraphics[width=\subplotwidth]{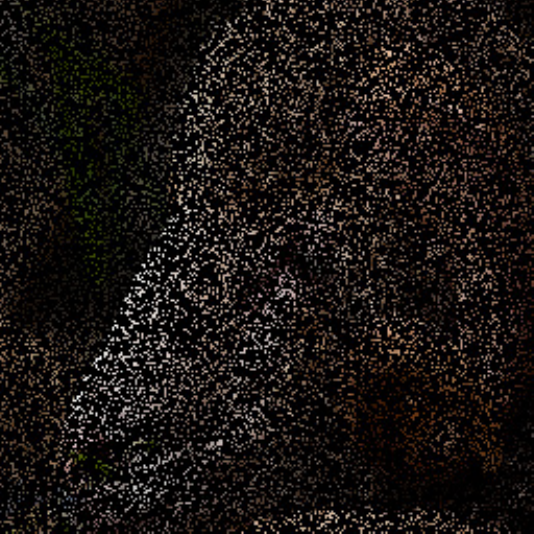}};
    \zoombox{0.58, 0.50}
\end{tikzpicture}
\begin{tikzpicture}[zoomboxarray]
    \node [image node] {\includegraphics[width=\subplotwidth]{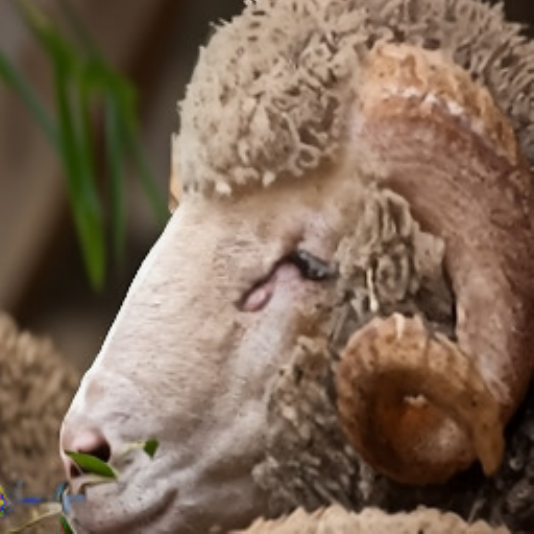}};
    \zoombox{0.58, 0.50}
\end{tikzpicture}
\begin{tikzpicture}[zoomboxarray]
    \node [image node] {\includegraphics[width=\subplotwidth]{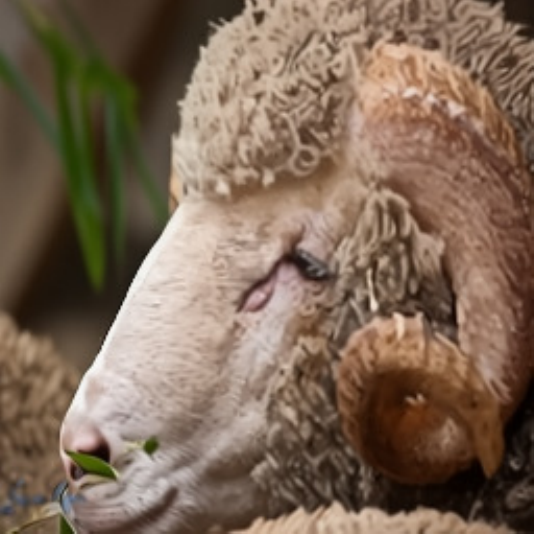}};
    \zoombox{0.58, 0.50}
\end{tikzpicture}
\begin{tikzpicture}[zoomboxarray]
    \node [image node] {\includegraphics[width=\subplotwidth]{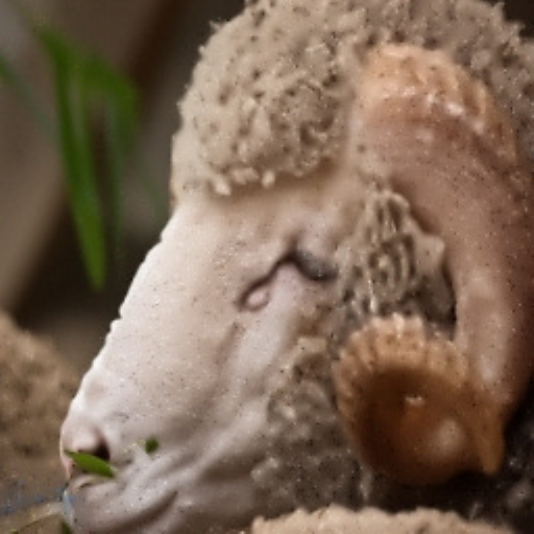}};
    \zoombox{0.58, 0.50}
\end{tikzpicture}
\begin{tikzpicture}[zoomboxarray]
    \node [image node] {\includegraphics[width=\subplotwidth]{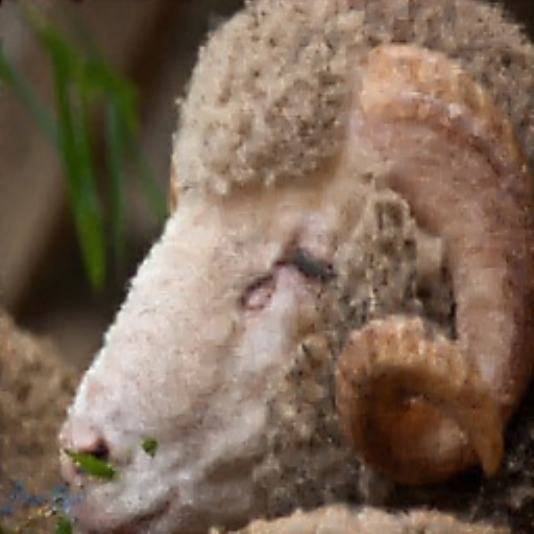}};
    \zoombox{0.58, 0.50}
\end{tikzpicture} 
    \\
\hspace{0.55cm}\begin{tikzpicture}[zoomboxarray]
    \node [image node] {\includegraphics[width=\subplotwidth]{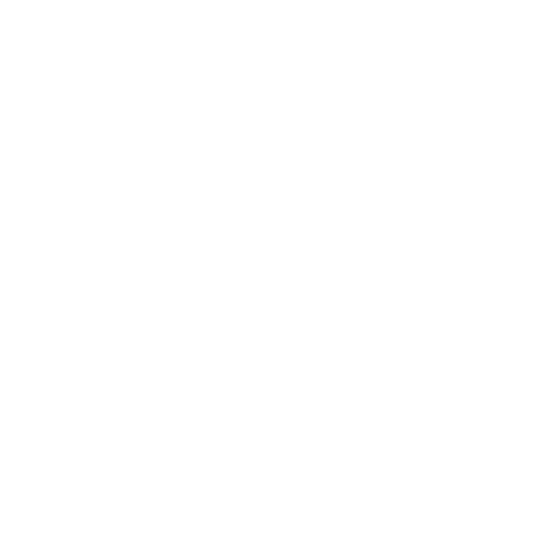}};
\end{tikzpicture} 
    \includegraphics[width=\subplotwidthbar]{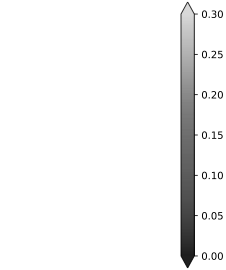}
\begin{tikzpicture}[zoomboxarray]
    \node [image node] {\includegraphics[width=\subplotwidth]{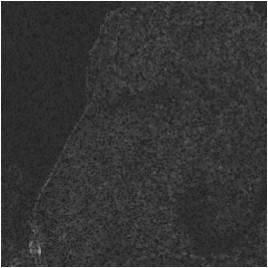}};
    \zoombox{0.58, 0.50}
\end{tikzpicture} 
\begin{tikzpicture}[zoomboxarray]
    \node [image node] {\includegraphics[width=\subplotwidth]{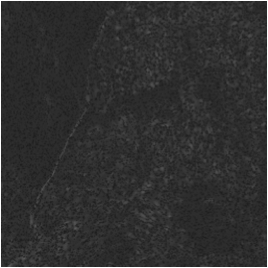}};
    \zoombox{0.58, 0.50}
\end{tikzpicture}
\begin{tikzpicture}[zoomboxarray]
    \node [image node] {\includegraphics[width=\subplotwidth]{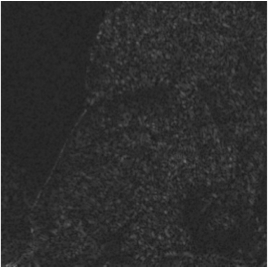}};
    \zoombox{0.58, 0.50}
\end{tikzpicture}
\begin{tikzpicture}[zoomboxarray]
    \node [image node] {\includegraphics[width=\subplotwidth]{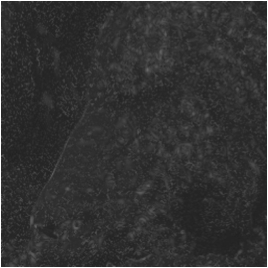}};
    \zoombox{0.58, 0.50}
\end{tikzpicture}
\caption{Inpainting task: images estimated by the compared algorithms (1st row) and corresponding uncertainty quantification (2nd row).}
\label{fig:imagenet}
\end{figure*}

\renewcommand{\subplotwidth}{.23\textwidth}
\begin{figure*}[ht!]
\centering
\begin{tabularx}{0.99\textwidth} { 
   >{\centering\arraybackslash}X 
   >{\centering\arraybackslash}X 
   >{\centering\arraybackslash}X 
   >{\centering\arraybackslash}X 
   }
\end{tabularx}
    \includegraphics[width=\subplotwidth]{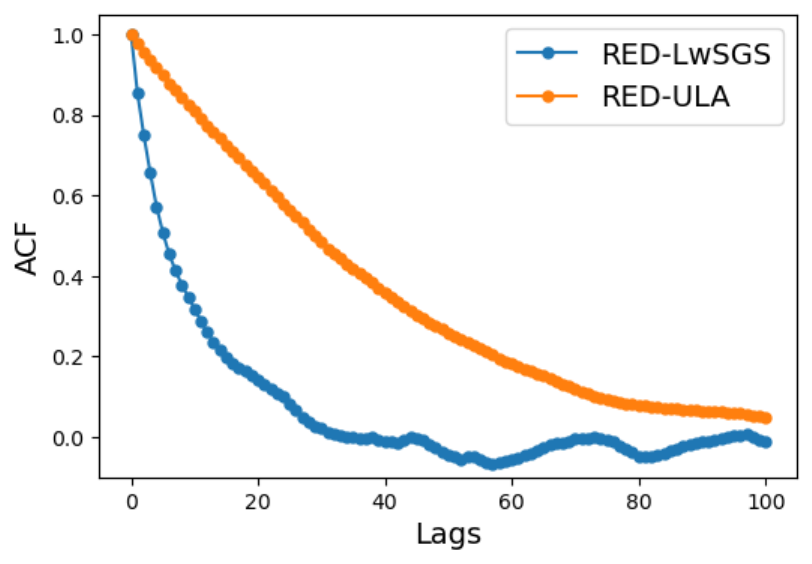}%
    \includegraphics[width=\subplotwidth]{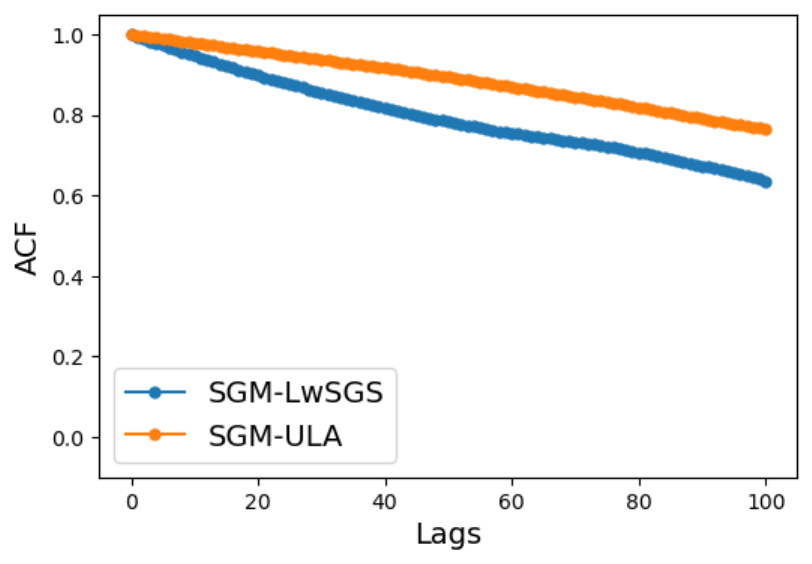}    	
    \includegraphics[width=\subplotwidth]{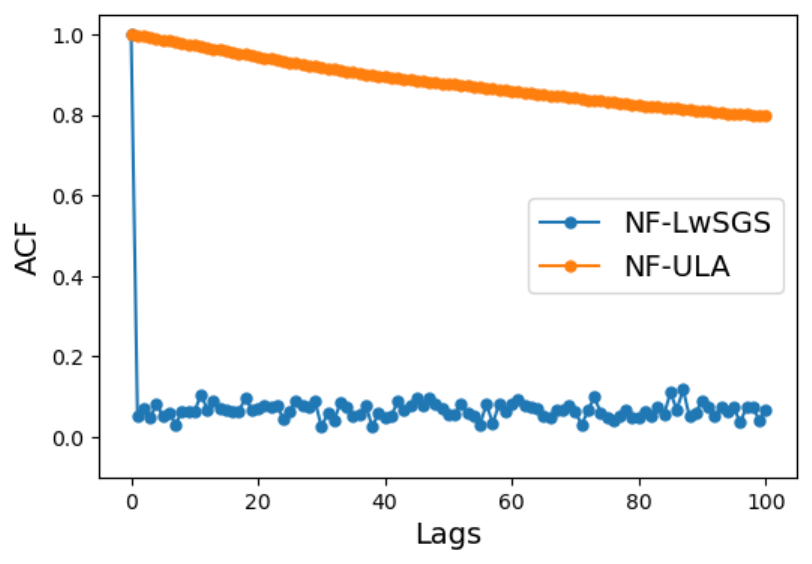}
    \includegraphics[width=\subplotwidth]{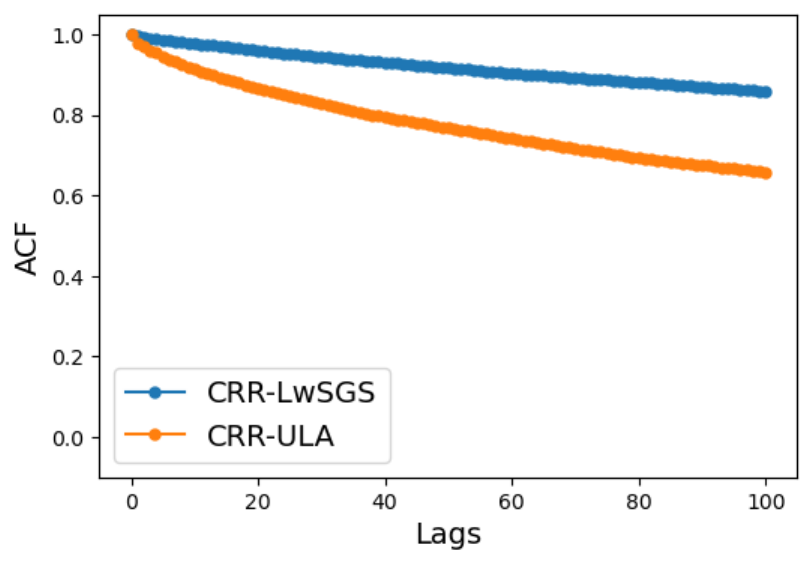}
    \\
    \includegraphics[width=\subplotwidth]{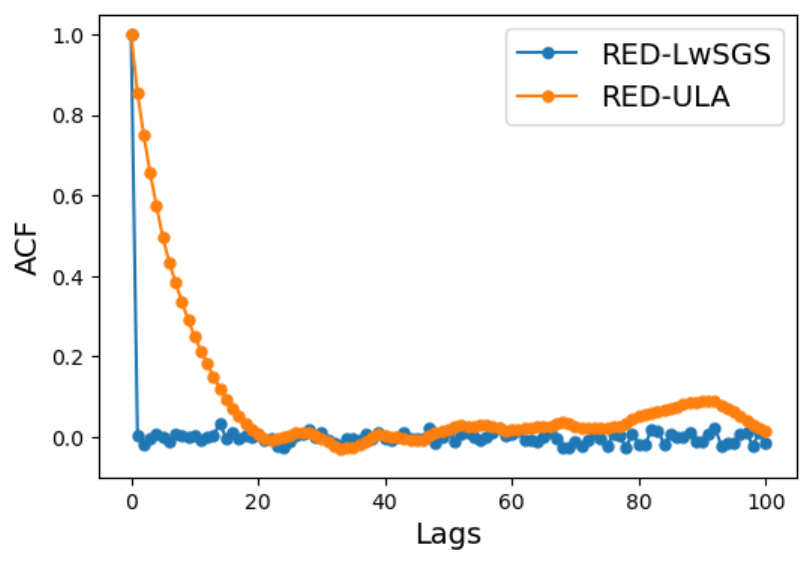}%
    \includegraphics[width=\subplotwidth]{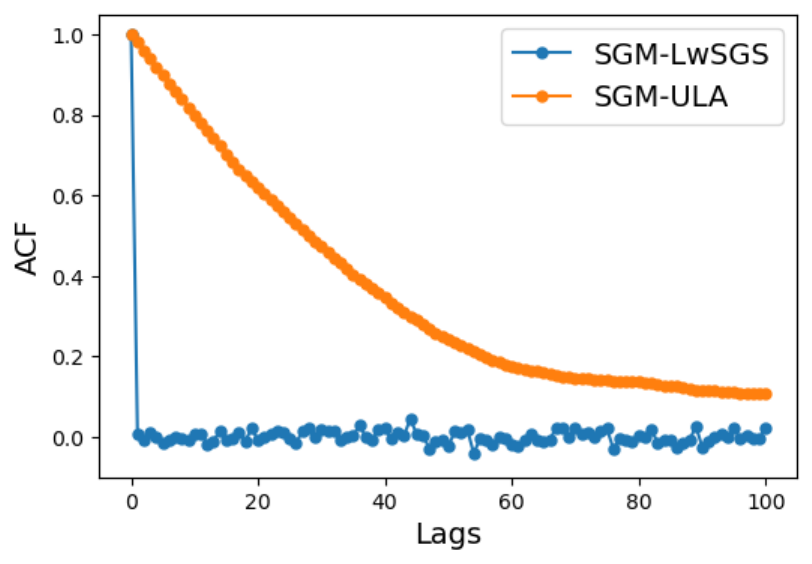}    	
    \includegraphics[width=\subplotwidth]{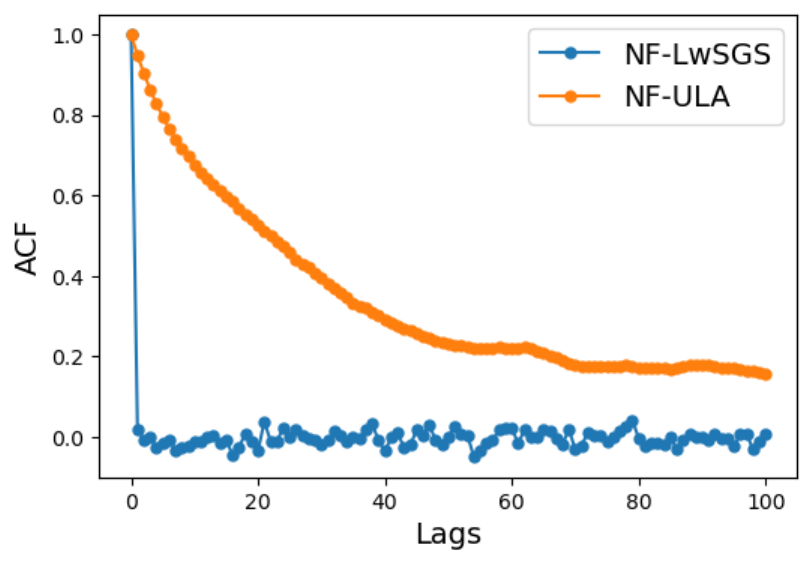}
    \includegraphics[width=\subplotwidth]{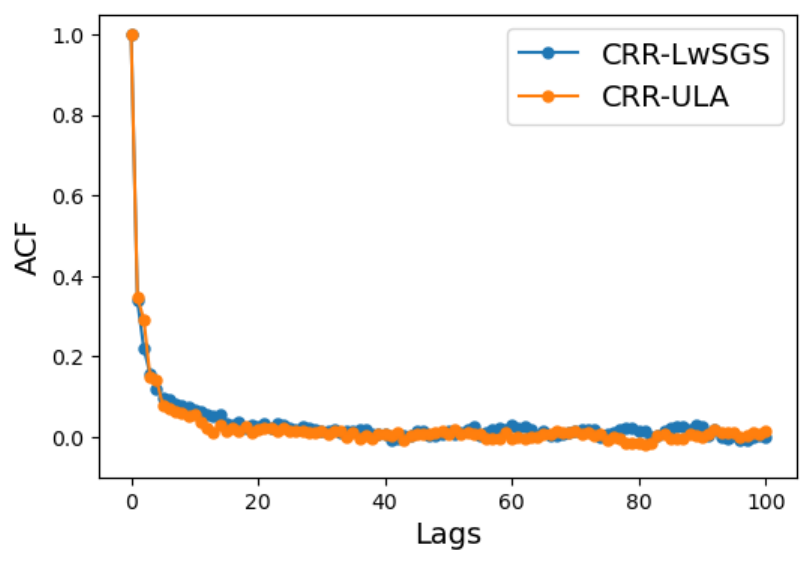}
     \\
\caption{Auto-correlation functions (ACF) of the generated samples (inpainting task, FFHQ data set): median direction (1st row) and fast direction (2nd row).}
\label{fig:afc_inpainting}
\end{figure*}

Table \ref{Tab:perfs_ffhq_imagenet}  also shows the computation times required by the compared algorithms. The computational time varies significantly across priors. It can be seen that LwSGS stands out for having the shortest computation times compared to the ULA framework.

The convergence properties of the compared sampling algorithms have been evaluated by examining the autocorrelation function (ACF) of the median and the fastest components of the chains generated by these algorithms. Denoting $\bx^{(t)} = \left[x_1^{(t)},\ldots,x_n^{(t)}\right]^{\top}$, the median and fastest components are defined as the produced pixel-wise Monte Carlo chain $\left\{x_{{i}}^{(t)}\right\}_{t=N_{\textrm{bi}}+1}^{N_{\mathrm{MC}}}$ with the median and largest variance, respectively. A faster decreasing ACF indicates less correlation between samples, which generally implies faster convergence of the Markov chain. Figure~\ref{fig:afc_inpainting} displays these ACFs for the inpainting task performed on an image from the FFHQ dataset. For the fast and median directions, the autocorrelation function (ACF) of the LwSGS framework exhibits a more rapid decrease compared to the ACFs obtained through the ULA framework across the majority of employed priors for the two considered tasks. This observation is supported by the IAT metric presented in Table \ref{Tab:perfs_ffhq_imagenet}.

\section{Application of RED-LwSGS on real data sets: restoring images of geological samples}\label{sec:real_data}

Identifying the nature of geomaterials (e.g., rocks, taillings, soils, sediments) is a key factor in geosciences for characterizing deposits, estimating resources and improving exploration, processing and environmental management operations. It provides information on rock weathering (color), petrophysical properties (permeability, porosity), surface state (roughness), texture (shape, size and spatial distribution of grains and minerals) and the presence of discontinuities and fractures \cite{alzubaidi2021automated,perez2018automated,xu2021deep}. This identification is classically carried out by geologists via visual inspection of samples that can result in a subjective interpretation of the nature of the geomaterials, in addition to being a long, laborious and costly task. To improve the reliability of this analysis, manufacturers are using analytical equipments such as scanners to acquire objective, reproducible RGB images at sample scale in a short space of time. These acquired RGB images are banked, providing an archive that can be consulted as often as required, and analyzed either by human operators or innovative data processing approaches. This section aims at investigating the performance of the proposed framework when restoring real-world imaging data of geological tailing samples acquired by the French Bureau of Geological and Mining Research (BRGM).

\subsection{Experimental setup and BRGM data sets}\label{subsec:BRGM_setup}

The image acquisition process employed a Basler acA2440-20gc camera mounted on a scanner to ensure stability during the capture process. The Basler acA2440-20gc, equipped with a Fujinon HF25HA-1S Lens (25 mm 1.5MP 2/3" f/1.4 C-Mount), captures RGB images of $1025 \times 2048$ pixels of tailings sample. To ensure consistent exposure and color of the images acquired, a uniform lighting composed of white LEB were used associated to a 18\% grey balance cards to set the camera illumination.

Using the previously described camera setup, real images (considered as ground-truth images) have been acquired in an optimally controlled manner. As an illustration, an in-focus image depicted in Fig. \ref{fig:real_data_ground} (left) clearly show that the tailing sample is composed of very micrometric whitish particles (clays) mixed with sandy black particles. These mixtures form centimetric aggregates separated from one another by voids corresponding to porosity and resulting in a rough surface. The  zoom carried out on the images makes it possible to distinguish individual particles, as well as discontinuities between the aggregates the edges of the sample. However, in real-world scenarios, it is common for acquired RGB images to show blurring over a portion or all of the sample, due to the presence of strong surface irregularities (height differences) and/or sensor limitations (e.g. depth of field, zoom), making the images unusable and consequently impossible to identify the nature of the geomaterials. To mimic such suboptimal acquisition protocols, out-of-focus images have been also acquired by playing on the focus to create blurred versions of the images (observations). On the corresponding out-of-focus image depicted in Fig. \ref{fig:real_data_ground} (right), it is now no longer possible to describe the relationships between the different particles that compose the aggregates, or to estimate their size. Similarly, estimating clear boundaries between aggregates or with sample edges is difficult. Without this information, it is not possible to describe the sample and therefore identify it, hence the importance of enhancing the visual quality of these images. The objective of the experiments conducted in this section aims at deblurring the out-of-focus BRGM images, i.e, recovering deblurred (i.e., sharper) images $\hat{\bx}\in \mathbb{R}^m$ from the real out-of-focus BRGM images $\by  \in \mathbb{R}^m$ assumed to result from the forward model \eqref{eq:forward_model}. 

\renewcommand{\subplotwidth}{.4\columnwidth}
\renewcommand{\subplotwidthbis}{.12\textwidth}
\renewcommand{\subplotwidthbisbis}{.05\textwidth}
\setlength{\tabcolsep}{1pt}
\begin{figure}[!h]
\centering
\begin{tabular}{ll}
    \begin{tikzpicture}[zoomboxarray]
    \node [image node] { \includegraphics[width=\subplotwidth]{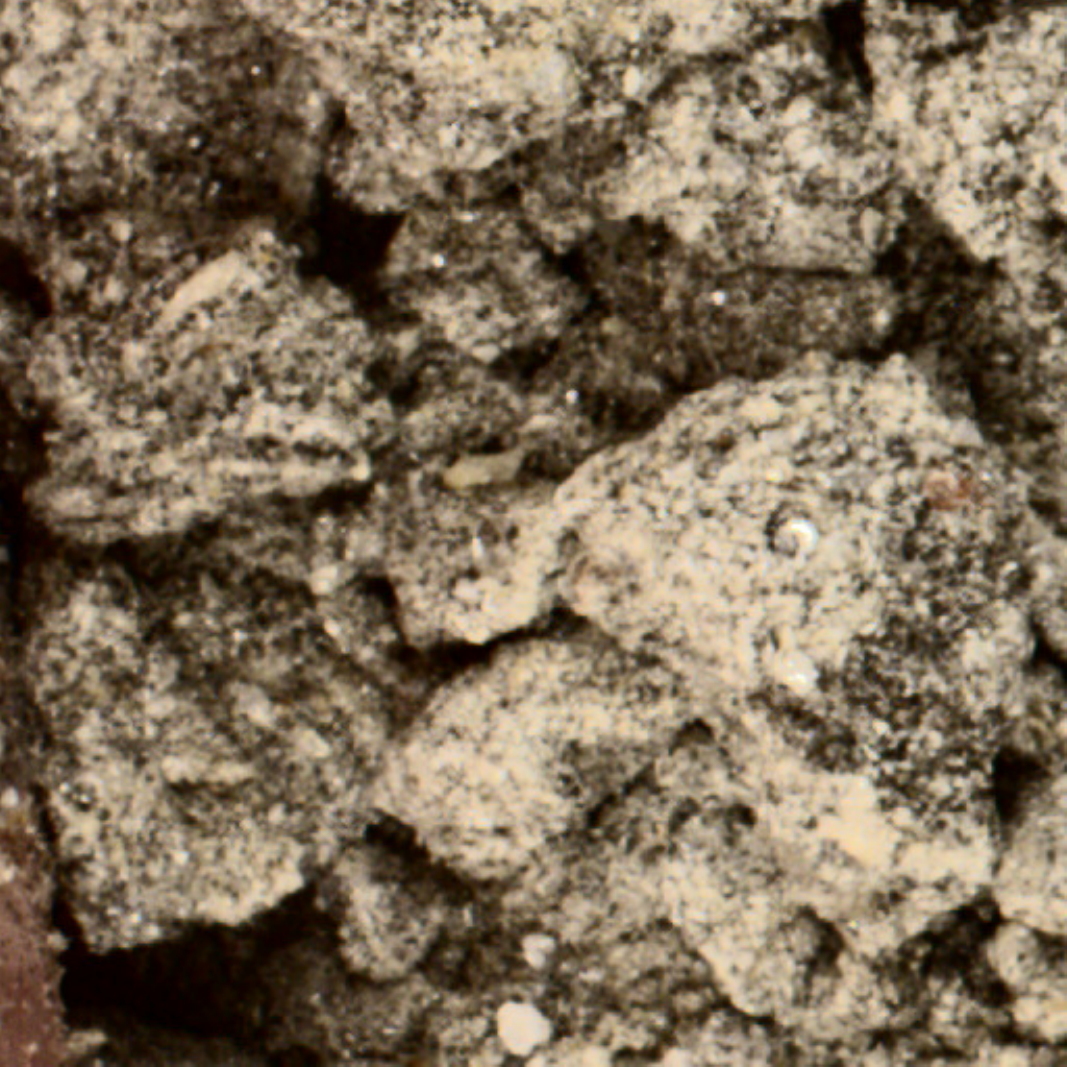}};
    \zoombox[magnification=2]{0.4, 0.6}
    \end{tikzpicture}
    &
    \begin{tikzpicture}[zoomboxarray]
    \node [image node] { \includegraphics[width=\subplotwidth,angle=180]{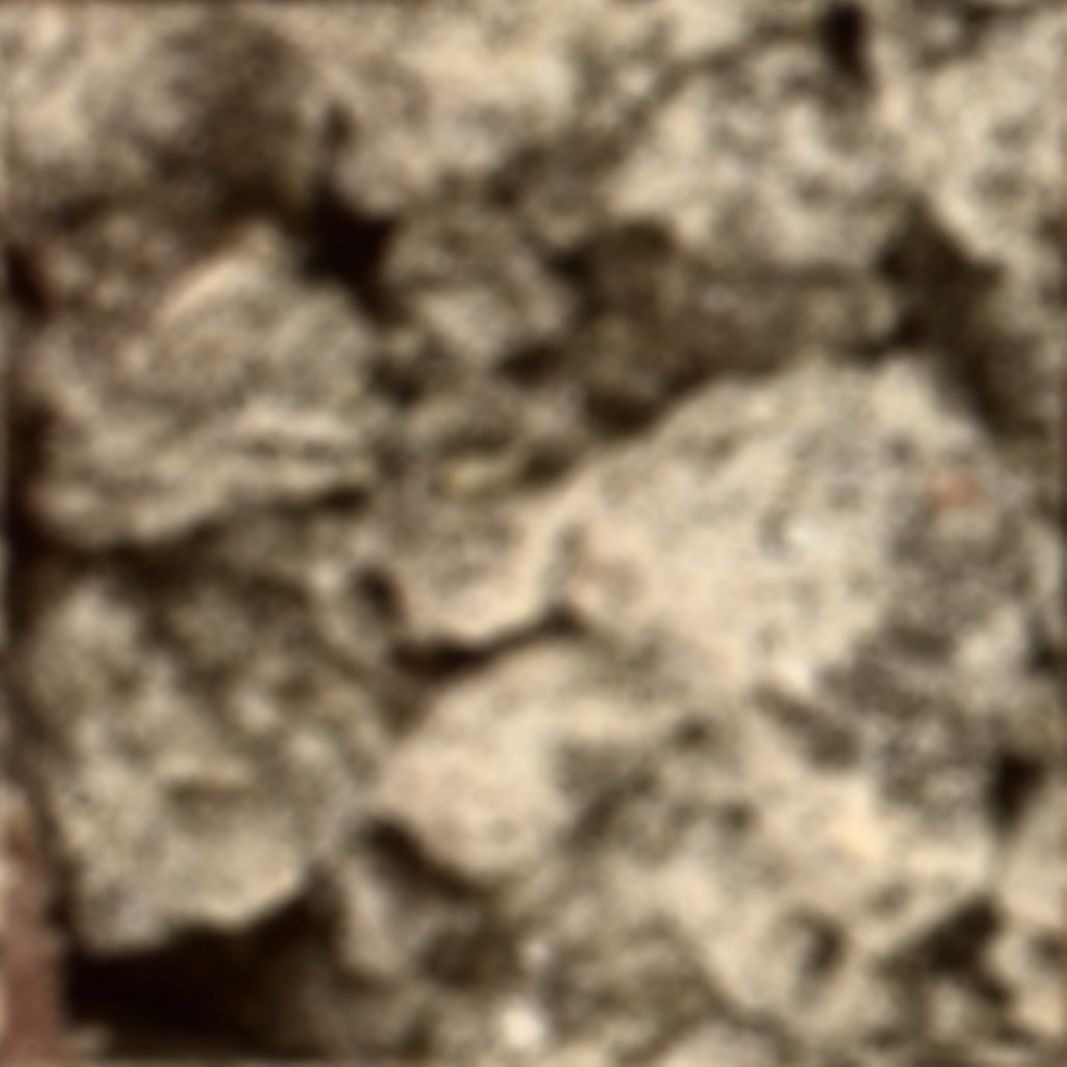}};
    \zoombox[magnification=2]{0.4, 0.6}
    \end{tikzpicture}
\end{tabular}
\caption{Real BRGM data set: in-focus (left) and out-of-focus (right) images.}
\label{fig:real_data_ground} 
\end{figure}

\subsection{Data preprocessing}

Before restoring the out-of-focus BRGM images acquired according to the protocol described in Section \ref{subsec:BRGM_setup}, a series of preprocessing steps are undertaken. First sub-images of size $512 \times 512$ pixels ($m=512^2$) are extracted from the original full images. This extraction is performed to ensure local consistency in both blur impact and restoration. The chosen sub-image size strikes a balance between computational efficiency and the accuracy of blur kernel estimation.

Additionally, to address variations in dynamic range across the images, histogram equalization is applied. Three distinct methods have been evaluated: equalization in individual RGB channels, equalization in the luminance channel of the YUV color space, and equalization in the value component of the HSV color space. Based on qualitative assessment, the YUV-based histogram equalization method has been selected, as it enhances contrast while maintaining color fidelity.

Finally, each pair of out-of-focus  and in-focus images is co-registered following a conventional pipeline of keypoint extraction \cite{alcantarilla2013} followed by homography estimation \cite{fischler1981random}.

\subsection{Blur kernel estimation}

In the considered experimental context of real acquisitions, the blurring operator $\bA \in \mathbb{R}^{m\times m}$ is unknown and should be estimated before restoration.  This operator is modeled as a 2D  spatially invariant linear filter fully parametrized by a blur kernel denoted as $\boldsymbol{\kappa}$, i.e.,  $\bA=\bA(\boldsymbol{\kappa})$. To estimate this blur kernel, one proposes to take advantage of the experimental protocol of  simultaneous acquisitions of both out-of-focus  and in-focus images. More precisely, let consider a set of $J$ blurred images $\bY_j$ where $\bY_j = \mathrm{unvec}(\by_j) \in \mathbb{R}^{\sqrt{m} \times \sqrt{m}}$ denotes the unvectorized representation of the image $\by_j$ ($j=1,\ldots,J$). According to  the acquisition setup described in Section \ref{subsec:BRGM_setup}, these of out-of-focus images are associated with $J$ corresponding co-registred in-focus images $\check{\bY}_j$ acquired over the same scene. Adopting the strategy detailed in \cite{pan2014deblurring}, the blur kernels are  estimated  by solving a least-square problem formulated in the gradient space
\begin{equation}
\operatornamewithlimits{min}_{\boldsymbol{\kappa}} \left\| \nabla \bY_j - \nabla \check{\bY}_j \ast \boldsymbol{\kappa} \right\|_2^2  + \lambda \left\|\boldsymbol{\kappa}\right\|^2_2 
\end{equation}
where $\ast$ stands for the 2D-convolution, $\nabla $ denotes the 2D discrete gradient operator (i.e, defined $1$st-order finite-differences in both directions) and the parameter $\lambda$ adjusts the amount of Tikhonov regularization. The degradation operator $\bA$ is finally computed after averaging over the kernels estimated on the $J$ image pairs.

\subsection{Image restoration results}

Once the degradation matrix $\bA$ has been estimated, the LwSGS algorithm is implemented when considering the RED-based prior. This choice of using this score function-based prior among the cases discussed in Section \ref{subsec:priors} is motivated by the findings reported in Section \ref{sec:experiments} which show that RED-based prior reaches the best trade-off between restoration performance and computational complexity. We recall that RED-LwSGS leverages the pre-trained DRUNet denoiser to define the prior distribution according to \eqref{RED_prior}. It generates samples asymptotically distributed according to the posterior distribution \eqref{split_dist}.   For this experiment, $N_{\text{MC}}=25\,000$ Monte Carlo iterations are used, including a burn-in period of $N_{\text{bi}}=10\,000$ iterations, to approximate the MMSE estimate of the restored image according to \eqref{eq:MMSE}. The regularization and coupling parameters are set to $\beta=5\times 10^{-3}$ and $\rho^2=9\times10^{-8}$, respectively. Experimental results obtained on two real BRGM images are depicted in Fig. \ref{fig:real_data}. For each image, this figure reproduces the blur kernel $\boldsymbol{\kappa}$ ($1$st row),  the observed out-of-focus image $\by$ ($2$nd row), the restored image $\hat{\bx}_{\mathrm{MMSE}}$ ($3$rd row) and the associated pixelwise standard deviations as  measures of uncertainty ($4$th row). Visually, these results show that the restored images exhibit significantly sharper details and illustrate the efficiency of the proposed method when restoring real images. This experimental validation opens the door towards the development of operational pipelines to exploit images of geological tailing samples that would have suffered from suboptimal acquisition protocols. 

\renewcommand{\subplotwidth}{.30\columnwidth}
\renewcommand{\subplotwidthbis}{.12\textwidth}
\renewcommand{\subplotwidthbisbis}{.05\textwidth}
\setlength{\tabcolsep}{1pt}
\begin{figure}[h!]
\centering
\begin{tabular}{ll}
    \begin{tikzpicture}[zoomboxarray]
    \node [image node] { \includegraphics[width=\subplotwidth]{estimated_kernel_1-eps-converted-to.pdf}};
    \end{tikzpicture}
    &
    \begin{tikzpicture}[zoomboxarray]
    \node [image node] { \includegraphics[width=\subplotwidth,angle=180]{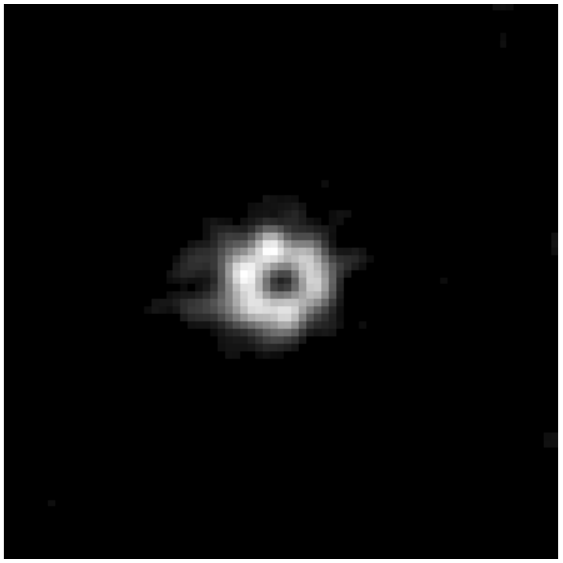}};
    \end{tikzpicture}
    \\
    \begin{tikzpicture}[zoomboxarray]
    \node [image node] { \includegraphics[width=\subplotwidth]{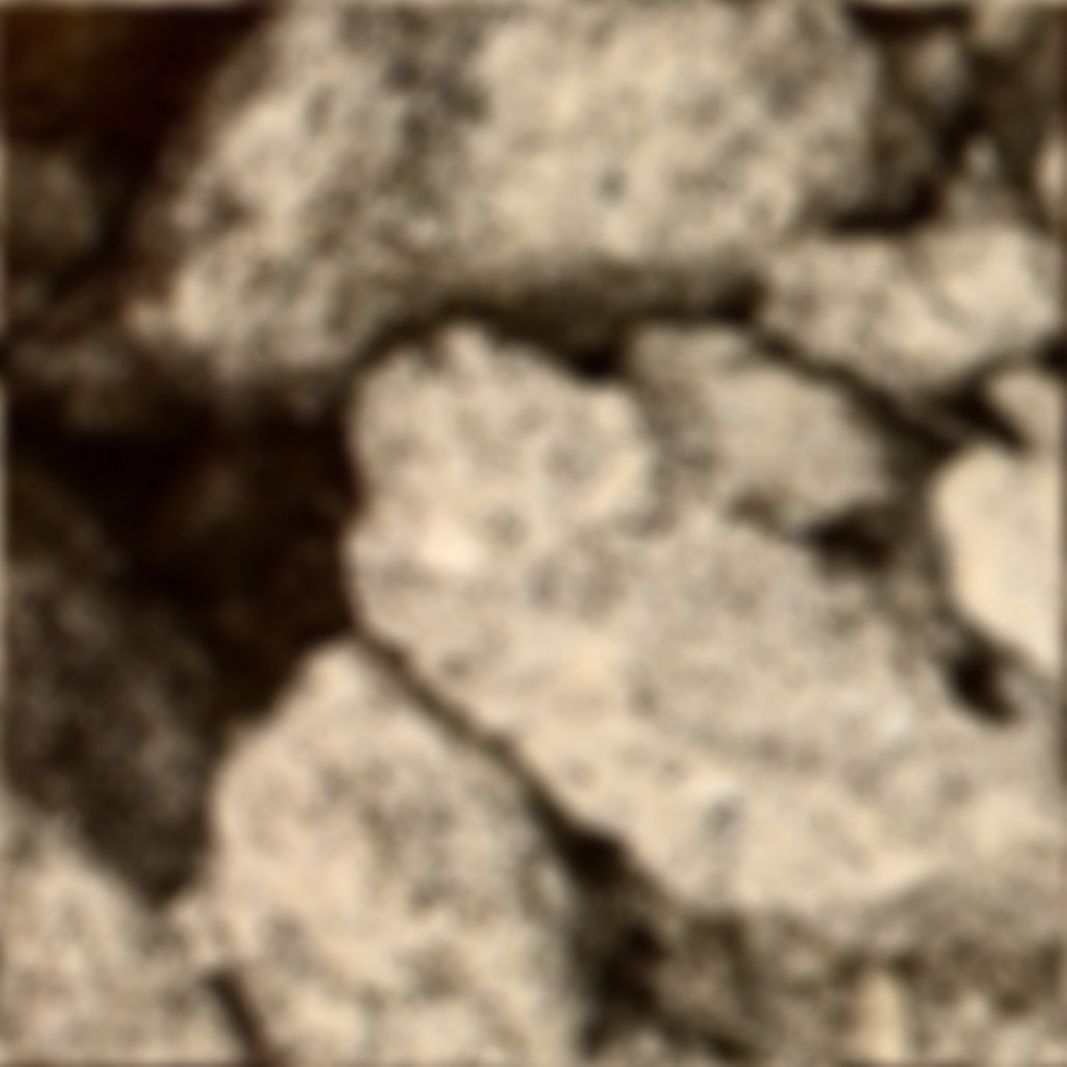}};
    \zoombox[magnification=2]{0.6, 0.65}
    \end{tikzpicture}
    &
    \begin{tikzpicture}[zoomboxarray]
    \node [image node] { \includegraphics[width=\subplotwidth,angle=180]{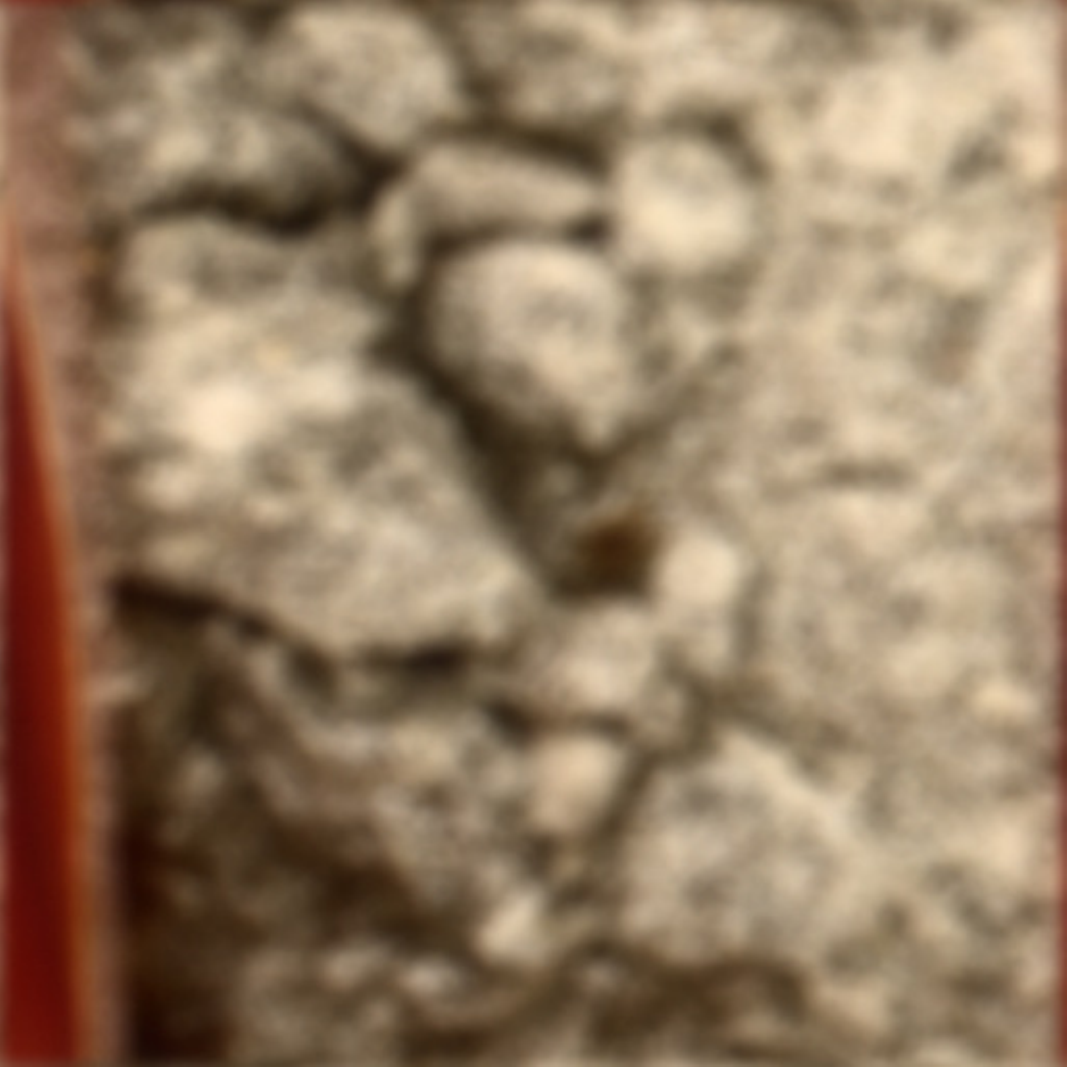}};
    \zoombox[magnification=2]{0.87, 0.65}
    \end{tikzpicture}
    \\
    \begin{tikzpicture}[zoomboxarray]
    \node [image node] { \includegraphics[width=\subplotwidth]{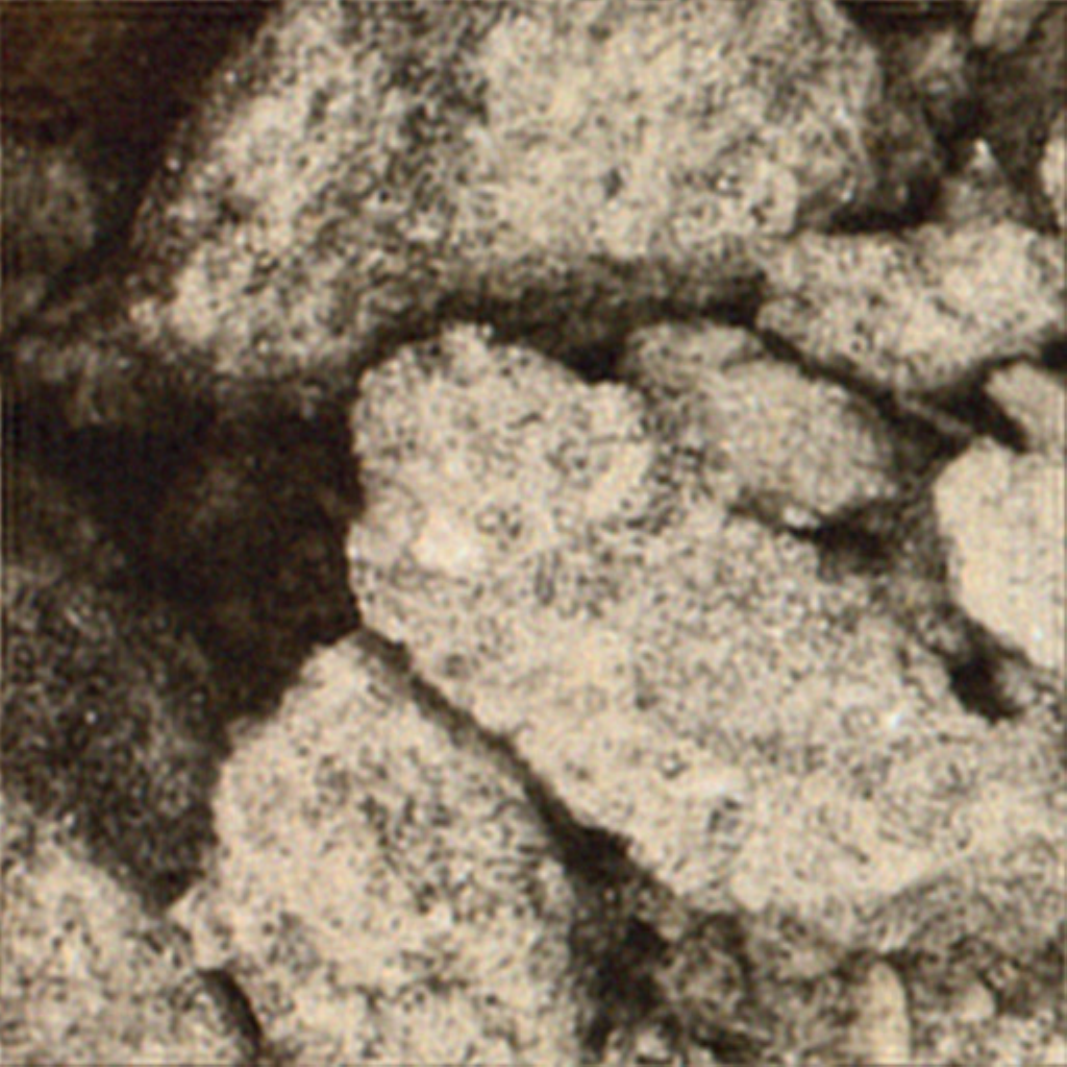}};
    \zoombox[magnification=2]{0.6, 0.65}
    \end{tikzpicture}
    &
    \begin{tikzpicture}[zoomboxarray]
    \node [image node] { \includegraphics[width=\subplotwidth,angle=180]{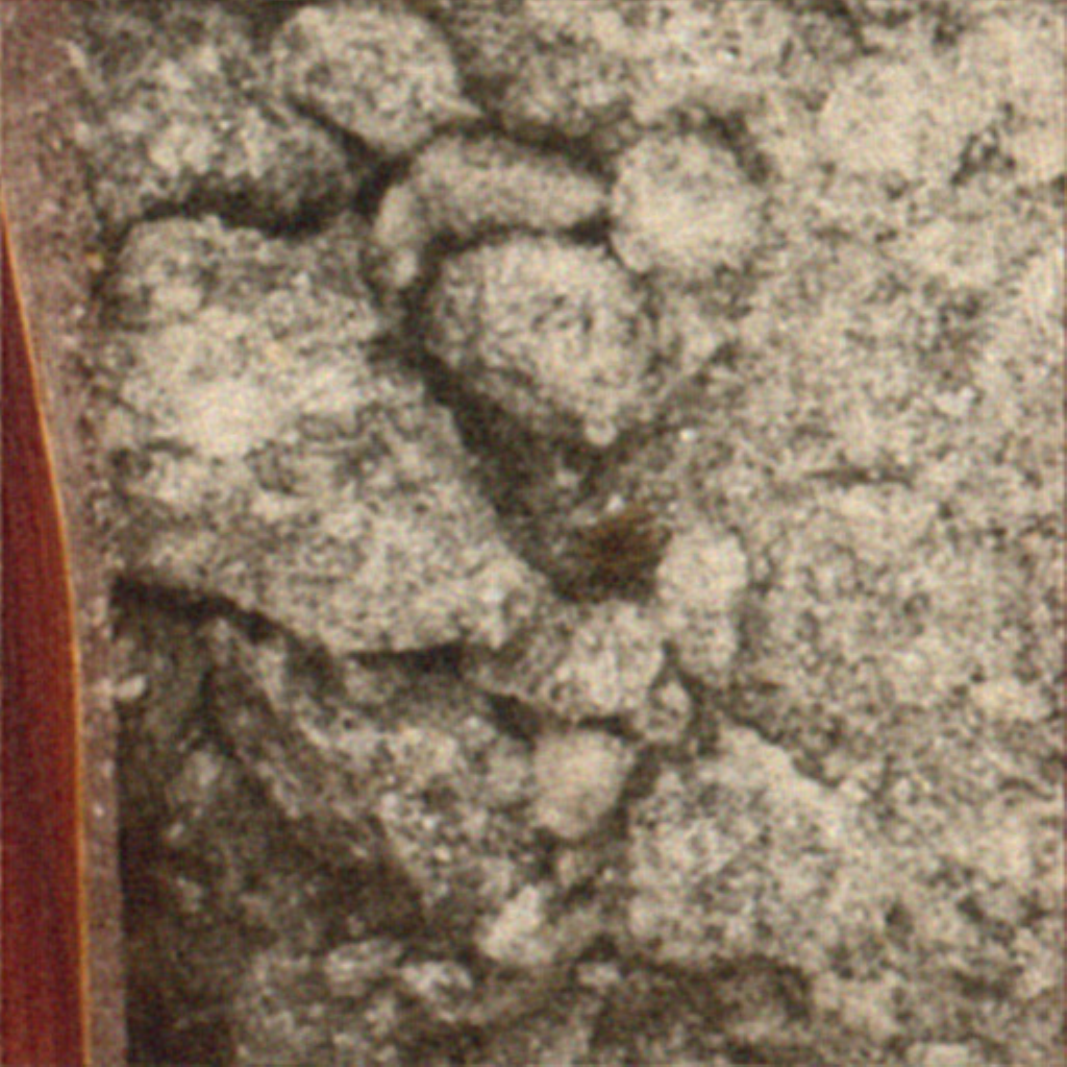}};
    \zoombox[magnification=2]{0.87, 0.65}
    \end{tikzpicture}
    \\
    \begin{tikzpicture}[zoomboxarray]
    \node [image node] { \includegraphics[width=\subplotwidth]{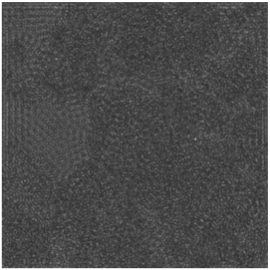}};
    \zoombox[magnification=2]{0.6, 0.65}
    \end{tikzpicture}  
    &
    \begin{tikzpicture}[zoomboxarray]
    \node [image node] { \includegraphics[width=\subplotwidth,angle=180]{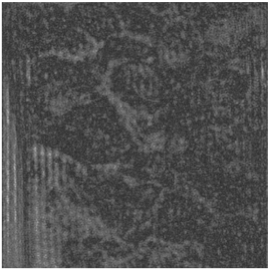}};
    \zoombox[magnification=2]{0.87, 0.65}
    \end{tikzpicture}   
\end{tabular}
\caption{Real BRGM data set: blur kernels ($1$st row), observed images ($2$nd row), restored images ($3$rd row), uncertainty quantifications ($4$th row).}
\label{fig:real_data}
\end{figure}

\section{Conclusion}
\label{sec:conclusion}
This paper showed that the Langevin-within-Split Gibbs Sampler (LwSGS) provided a systematic and versatile alternative to the unadjusted Langevin algorithm (ULA) to sample from posterior distributions characterized by score function-based priors which were explicit or easily computable. By employing an asymptotically exact data augmentation (AXDA), LwSGS effectively separated the data-fitting and regularization terms into two conditional posterior distributions from which it was easier to sample. To sample from the conditional distribution which embeds the prior model, LwSGS adopted a Langevin Monte Carlo step to benefit from explicit form of the prior score function. The methodology was demonstrated to accommodate diverse prior distributions defined by score functions, namely regularization-by-denoising, score-based generative models, normalizing flows and convex ridge regularizations. Numerical experiments illustrated that this approach compares favorably with ULA-based sampling schemes. The application of the proposed framework to the restoration of real data acquired in geological context represented a hope to exploit images of geological residue samples that would had suffered from suboptimal acquisition protocols. This unified approach not only offers a robust and flexible solution for Bayesian inverse problems but also paves the way for the integration of diverse, score function-based priors beyond the specific cases considered here. Future works include investigating theoretical guarantees about the convergence of the LwSGS algorithm when sampling from posterior distributions defined by score function-based priors, extending the results in \cite{Faye2024}.
\bibliographystyle{elsarticle-num}
\bibliography{strings_all_ref,biblio}

\end{document}